\documentclass[twocolumn,superscriptaddress]{revtex4-2}
\usepackage[utf8]{inputenc}
\usepackage{xcolor}
\usepackage{braket}  % allows Braket notation
\usepackage{graphics,graphicx,epsfig}
\usepackage{amssymb,amsfonts,amsmath}
\usepackage{ifthen}
\usepackage[pdftex]{hyperref}
\usepackage{hyperref}
\graphicspath{{Figures/}}		
\usepackage{enumitem}
\newcommand{\EQ}{\begin{equation}}
\newcommand{\EE}{\end{equation}}
\newcommand{\EQA}{\begin{eqnarray}}
\newcommand{\EEA}{\end{eqnarray}}

\newcommand{\eff}{{\text{eff}}}
\newcommand{\attr}{{\text{att}}}
\newcommand{\lag}{{{}_{\text{lag}}}}
\newcommand{\att}{{\text{att}}}

\newcommand{\Q}{{\mathcal{Q}}}
\newcommand{\E}{\mathbb{E}}

\newcommand{\rH}{\mathrm{H}}

\newcommand{\rS}{\mathrm{S}}
\newcommand{\A}{\mathcal{A}}
\newcommand{\MI}{{\text{MI}}}
\DeclareMathOperator*{\argmax}{argmax}
\DeclareMathOperator*{\argmin}{argmin}
\usepackage{hyperref}
\newcommand{\av}[1]{\E\left[{#1}\right]}

\newcommand{\C}{\mathrm{c}}

\usepackage[normalem]{ulem}
\usepackage{dsfont}

\begin{document}
		
\title{Learning and organization of memory for evolving patterns}
\author{Oskar H Schnaack}
\address{Max Planck Institute for Dynamics and Self-organization, Am Fa\ss berg 17, 37077 G\"ottingen, Germany}
\address{Department of Physics, University of Washington, 3910 15th Ave Northeast, Seattle, WA 98195, USA}
\author{Luca Peliti}
\address{Santa Marinella Research Institute, 00058 Santa Marinella, Italy}
\author{Armita Nourmohammad}\thanks{Correspondence should be addressed to: armita@uw.edu. }
\address{Max Planck Institute for Dynamics and Self-organization, Am Fa\ss berg 17, 37077 G\"ottingen, Germany}
\address{Department of Physics, University of Washington, 3910 15th Ave Northeast, Seattle, WA 98195, USA}
\address{Fred Hutchinson Cancer Research Center, 1100 Fairview ave N, Seattle, WA 98109, USA}

\date{\today} 
\begin{abstract}
\noindent 
Storing memory for molecular recognition is an efficient strategy for responding to external stimuli. Biological processes use different strategies to store memory. In the olfactory cortex, synaptic connections form when stimulated by an odor, and establish distributed memory that can be retrieved upon re-exposure. In contrast, the immune system encodes specialized memory by diverse receptors that recognize a multitude of  evolving pathogens. Despite the mechanistic differences between the olfactory and the immune memory, these systems can still be viewed as different information encoding strategies. Here, we present a theoretical framework with artificial neural networks to characterize optimal memory strategies for both static and dynamic (evolving) patterns. Our approach is a generalization of the energy-based Hopfield model in which memory is stored as a network's energy minima. We find that while classical Hopfield networks with distributed memory can efficiently encode a memory of static patterns, they are inadequate against evolving patterns. To follow an evolving pattern, we show that a distributed network should use a higher learning rate, which in turn, can distort the energy landscape associated with the stored memory attractors. Specifically, narrow connecting paths emerge between memory attractors, leading to misclassification of evolving patterns. We demonstrate that compartmentalized networks with specialized subnetworks are the optimal solutions to memory storage for evolving patterns. We postulate that evolution of pathogens may be the reason for the immune system to encoded a focused memory, in contrast to the distributed memory used in the olfactory cortex that interacts with mixtures of static odors. 
\end{abstract}
\keywords{ }
\maketitle

\section{Introduction}

Storing memory for molecular recognition is an efficient strategy for sensing and response to external stimuli. Apart from the cortical memory in the nervous system, molecular memory is also an integral part of the immune response, present in a broad range of organisms from the CRISPR-Cas system in bacteria~\cite{Labrie:2010bc,Barrangou:2014ht,Bradde:2020kb} to adaptive immunity in vertebrates~\cite{Perelson:1997dla,Janeway:2001te,AltanBonnet:2020hk}. In all of these systems, a molecular encounter is encoded as a memory and is later retrieved and activated in response to a similar stimulus, be it a  pathogenic  reinfection or a re-exposure to a pheromone. Despite this high-level similarity, the immune system and the  synaptic nervous system utilize vastly distinct molecular mechanisms for storage and retrieval of their  memory.

Memory storage, and in particular, associative memory  in the hippocampus and olfactory cortex has been a focus of theoretical and computational studies in neuroscience~\cite{Haberly:1989jo,Brennan:1990bp,Granger:1991hu,Haberly:2001gx,Wilson:2004do}. In the case of the olfactory cortex, the input is a combinatorial pattern produced by olfactory receptors which recognize the constituent  mono-molecules of a given odor.  A given odor composed of many mono-molecules at different concentrations \cite{Raguso:2008re,Dunkel:2014na,Beyaert:14br} drawn from a space of about $10^4$ distinct mono-molecules reacts with olfactory receptors (a total of $\sim 300-1000$ in mammals \cite{glusman:2001gr,Bargmann:2006na,Touhara:2009AP,su:2009cl,verbeurgt:2014po}), resulting in a distributed signal and spatiotemporal pattern in the olfactory bulb~\cite{Shepherd:1998ur}.  This pattern is transmitted to the olfactory cortex, which serves as a pattern recognition device and enables an organism to distinguish orders of magnitudes more odors compared to the number of olfactory receptors \cite{Bushdid:2014sc,Gerkin:2015el,Mayhew:2020bx}. The synaptic connections in the cortex are formed as they are co-stimulated by a given odor pattern, thus forming an associative memory that can be retrieved in future exposures~\cite{Haberly:1989jo,Brennan:1990bp,Granger:1991hu,Haberly:2001gx,Wilson:2004do,Lansner:2009hb}.

\begin{figure*}[]
\centering
\includegraphics[width= \textwidth]{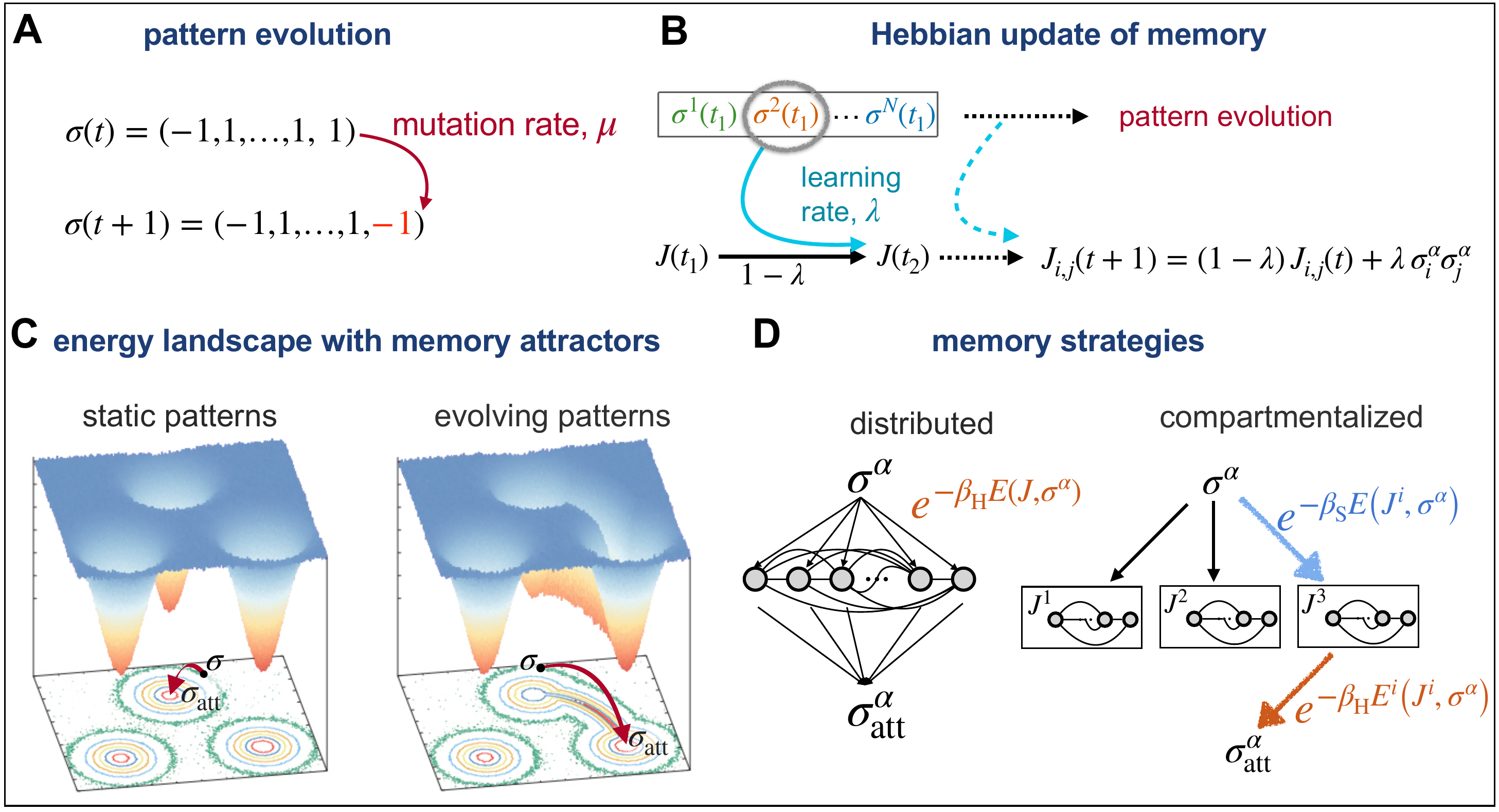}
 \caption{ {\bf Model of Working memory for evolving patterns.} {\bf (A)} The targets of recognition are encoded by binary vectors $\{\sigma\}$ of length $L$. Patterns can evolve over time with a mutation rate $\mu$, denoting  the fraction of spin-flips in a pattern per network update event. {\bf (B)}  Hebbian learning rule is shown for network $J$, which is presented a set of $N$ patterns $\{\sigma^\alpha\}$ (colors) over time. At each step, one pattern $\sigma^\alpha$ is randomly presented to the network and the network is updated with learning rate $\lambda$ (eq.~\ref{eq:update}).  {\bf(C)} The energy landscape for networks with distributed memory with optimal learning rate for static (left) and evolving (right) patterns are shown. The  equipotential lines are shown in the bottom 2D plane. The energy minima correspond to memory attractors. For static patterns (left), equilibration in the network's energy landscape drives a patterns towards its associated memory attractor, resulting in an accurate reconstruction of the pattern.  For evolving patterns (right), the heightened optimal learning rate of the network results in the emergence of  connecting paths (mountain passes) between the energy minima. The equilibration process can drive a pattern through a mountain pass towards a wrong memory attractors, resulting in pattern misclassification.  {\bf (D)}  A network with distributed memory (left) is compared to a specialized network with multiple compartments (right). To find an associative  memory, a presented pattern $\sigma^\alpha$ with energy $E(J,\sigma^\alpha)$ in network $J$ equilibrates with inverse temperature $\beta_\rH$ in the network's energy landscape and falls into an energy attractor $\sigma^\alpha_\attr$. Memory retrieval is a two-step process in a compartmentalized network (right): First, the sub-network $J^i$ is chosen with a probability $P_i\sim \exp[-\beta_\rS E^i(J^i,\sigma^\alpha)]$, where  $\beta_\rS$ is the inverse temperature for this decision. Second, the pattern equilibrates within the sub-network and falls into an energy attractor $\sigma^\alpha_\attr$.  }
\label{fig:Fig1} 
\end{figure*}

Artificial neural networks that store auto-associative memory are used to model the olfactory cortex. In these networks, input patterns   trigger interactions between encoding nodes. The ensemble of interactive nodes keeps a robust memory of the pattern since they can  be  simultaneously stimulated upon re-exposure and thus a stored pattern can be recovered  by just knowing part of its content. This mode of memory storage resembles the co-activation of synaptic connections in a cortex. Energy-based models, such as Hopfield neural networks with Hebbian update rules~\cite{Hebb:1949vs}, are among such systems, in which memory is stored as the network's energy minima~\cite{Hopfield:1982fq}; see Fig.~\ref{fig:Fig1}. The connection between the standard Hopfield network and the real synaptic neural networks has been a subject of debate over the past decades. Still, the Hopfield network provides a simple and solvable coarse-grained model of the synaptic network, relevant for working memory in the olfactory system and the hippocampus~\cite{Lansner:2009hb}.

Immune memory is encoded very differently from  associative memory in the nervous system and the olfactory cortex. First, unlike olfactory receptors,  immune receptors are extremely diverse and can specifically recognize pathogenic molecules without the need for a distributed and combinatorial encoding. In vertebrates, for example, the adaptive immune system consists of tens of billions of  diverse B- and T-cells, whose unique surface receptors are generated through genomic rearrangement, mutation, and selection, and can recognize and mount specific responses against the multitude of pathogens~\cite{Janeway:2001te}. Immune cells activated in response to a pathogen can differentiate into memory cells, which are long lived and can more efficiently  respond upon reinfections. As in most molecular interactions, immune-pathogen recognition is cross-reactive, which would allow memory receptors to recognize slightly evolved forms of the pathogen~\cite{Janeway:2001te,Mayer:2015ce,Shinnakasu:2016ei,Shinnakasu:2017ct,Mayer:2019is,Schnaack:2020vb,Viant:2020jt}. Nonetheless, unlike the distributed memory in the olfactory cortex, the receptors encoding immune memory  are focused and can only interact with pathogens with limited evolutionary divergence from the primary infection, in response to which memory was originally generated~\cite{Janeway:2001te}.

There is no question that there are vast mechanistic and molecular differences between how memory is stored in the olfactory system compared to the immune system. However, we can still ask which key features of the two systems prompt such distinct information encoding strategies for their respective distributed versus specialized  memory. To probe the emergence of distinct memory strategies, we propose a generalized Hopfield model that can learn and store memory against both the static and the dynamic (evolving) patterns. We formulate this problem as an optimization task to find a  strategy (i.e., learning rate and network structure) that confer the highest accuracy for memory retrieval in a network (Fig.~\ref{fig:Fig1}). 

In contrast to the static case, we show that a distributed memory in the style of a classical Hopfield model~\cite{Hopfield:1982fq} fails to efficiently work for evolving patterns. We show that the optimal learning rate should increase with faster  evolution of patterns, so that a network can follow the dynamics of the evolving patterns. This heightened learning rate tends to carve narrow connecting paths (mountain passes) between the memory attractors of a network's energy landscape, through which patterns can equilibrate in and be associated with a wrong memory. To overcome this misclassification, we demonstrate that specialized memory  compartments emerge in a neural network  as an optimal solution to efficiently recognize and retrieve a memory of out-of-equilibrium evolving patterns. Our results suggest that evolution of pathogenic patterns may be one of the key reasons why the immune system encodes a focused memory, as opposed to the  distributed memory used in the olfactory system, for which molecular mixtures largely present  static patterns. Beyond this biological intuition, our model offers a principle-based analytical framework to study learning and memory generation in out-of-equilibrium dynamical systems.

%%------------------------
\section{Results}
\subsection{Model of working memory for evolving patterns}

\begin{figure*}[t!]
\includegraphics[width=\textwidth]{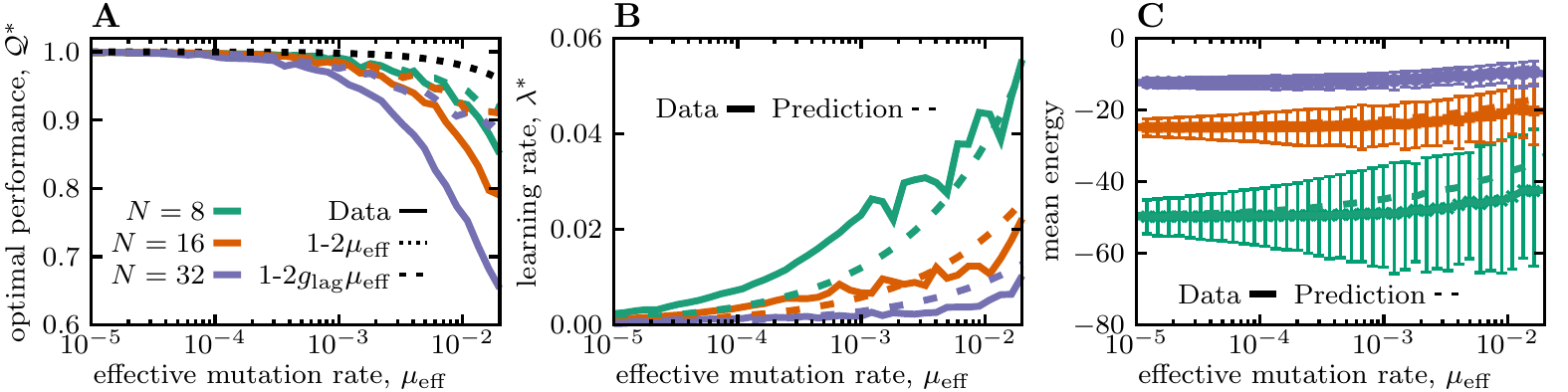}
\caption{\label{Fig2}
{\bf Reduced performance of Hopfield networks in retrieving memory of evolving patterns.} {\bf (A)} The optimal performance of a network $\Q^*\equiv \Q(\lambda^*)$ (eq.~\ref{eq.accuracy}) is shown as a function of the effective mutation rate $\mu_{\eff}= N\mu$. The solid lines show the simulation results for networks encountering different number of patterns (colors). The black dotted line shows the na\"ive expectation for the performance solely based on the evolutionary divergence of the patterns $\Q_0 \approx 1 - 2 \mu_{\eff}$, and the colored dashed lines show the expected performance after accounting for the memory lag $g_\lag$, $\Q_\text{lag} \approx 1 - 2g_\lag \mu_\eff$; see Fig.~\ref{fig:Simresults_overleps_high} for more details.  
{\bf (B)} The optimal learning rate $\lambda^*$ is shown as a function of effective mutation rate.  The solid lines are the numerical estimates and dashed lines show the theoretical predictions (eq.~\ref{eq:lambda}). {\bf (C)} The mean energy obtained by simulations of randomly ordered patterns (solid lines) and the analytical approximation (eq.~\ref{eq:expected_energy_order}) for ordered patterns (dotted lines) are shown. Error bars show standard error from  the independent realizations (Appendix A). The color code  for the number of presented patterns is consistent across panels, and the length of patterns is set to  $L=800$.}
\end{figure*}

To probe memory strategies against different types of stimuli, we propose a generalized energy-based model of associative memory, in which a Hopfield-like neural network is able to learn and subsequently recognize binary patterns. This neural network is characterized by an energy landscape and memory is stored as the network's energy minima.  We encode the target of recognition (stimulus) in a binary vector~$\sigma$ (\textit{pattern}) with $L$ entries: $\sigma=(\sigma_1,\ldots,\sigma_L)$, with $\sigma_i=\pm 1$, $\forall i$ (Fig.~\ref{fig:Fig1}A).  To store associative memory, we define a fully connected network represented by an interaction matrix~$J=(J_{i,j})$ of size~$L \times L$, and use a Hopfield-like energy function (Hamiltonian) to describe pattern recognition ${E_J(\sigma)=-\frac{1}{2L}\sum_{ij}J_{i,j}\sigma_i\sigma_j \equiv -\frac{1}{2}\braket{\sigma|J|\sigma}}$~\cite{Hopfield:1982fq} (Fig.~\ref{fig:Fig1}C). Here, we used a short-hand notation to denote the normalized pattern vector by $\ket\sigma \equiv \frac{1}{\sqrt{L}} \sigma$, its transpose by $\bra\sigma$, resulting in a  normalized scalar product  ${\braket{\sigma|\sigma'}\equiv \frac{1}{L}\sum_i\sigma_i\sigma'_i}$, and a matrix product ${\braket{\sigma| J|\sigma} \equiv\frac{1}{L} \sum_{i,j} \sigma_i J_{i,j}\sigma_j}$.

The network undergoes a learning process, during which different patterns are presented sequentially and  in random  order (Fig.~\ref{fig:Fig1}B). As a pattern $\sigma^\alpha$ is presented, the interaction matrix~$J$ is updated according to the following Hebbian update rule~\cite{workingMem87}
\begin{equation}
J_{i,j}\longrightarrow J'_{i,j}=
\begin{cases}
(1-\lambda)\,J_{i,j}+\lambda \,\sigma^{\alpha}_i\sigma^{\alpha}_j,&\text{if }i\neq j;\\
0,&\text{otherwise.}
\end{cases}
\label{eq:update}
\end{equation}
Here $\lambda$ is the \textit{learning rate}. In this model, the memorized patterns are represented by energy minima associated with the matrix $J$. We consider the case where the number $N$ of different pattern classes is below the {\em Hopfield capacity} of the network (i.e., $N\lesssim 0.14\, L$; see refs.~\cite{Hopfield:1982fq,Amit:1985bo,McEliece:1987ie}).

With the update rule in eq.~\ref{eq:update}, the network develops energy minima as associative memory close to each of the  previously presented  pattern classes $\sigma^\alpha$ ($\alpha\in\{1,\ldots,N\}$)(Fig.~\ref{fig:Fig1}C). Although the network also has minima close to the negated patterns, i.e., to~$-\sigma^\alpha$, they do not play any role in what follows.  
To find an associative memory we let  a presented pattern $\sigma^\alpha$ equilibrate in the  energy landscape, whereby we accept spin-flips $\sigma^\alpha \to \tilde\sigma^\alpha$ with a probability $\min\left(1,e^{-\beta_\rH \left(E_J(\tilde\sigma) - E_J(\sigma) \right) } \right)$, where  $\beta_\rH$ is the inverse equilibration (Hopfield) temperature (Appendix A). In the low temperature regime (i.e., high $\beta_\rH$), equilibration in networks with working memory  drives a presented pattern $\sigma^\alpha$ towards a similar attractor $\sigma^\alpha_\attr$, reflecting the memory associated with the corresponding energy minimum (Fig.~\ref{fig:Fig1}C).  This similarity is measured by the {\em overlap} $q^\alpha \equiv \braket{\sigma^\alpha_\attr|\sigma^\alpha}$ and determines the accuracy of the associative memory. 

Unlike the classical cases of pattern recognition by Hopfield networks, we assume that patterns can evolve over time with a rate $\mu$ that reflects the average number of spin-flips in a given pattern per network's update event (Fig.~\ref{fig:Fig1}A). Therefore, the expected number of spin-flips  in a given pattern between two encounters is  $\mu_{\eff}=N\mu$, as two successive encounters of the same pattern are on average separated by $N-1$ encounters (and updates) of the network  with the other patterns. We work in the regime where the mutation rate~$\mu$ is  small enough such that the evolved patterns stemming from a common ancestor $\sigma^\alpha(t_0)$ at time $t_0$  (i.e., the members of the \textit{class} $\alpha$) remain more similar to each other than to members of the other classes (i.e., $\mu N L \ll L/2$).  

The special case of  static patterns ($\mu_\eff=0$) can reflect distinct odor molecules, for which associative memory is stored in the olfactory cortex.  On the other hand, the distinct pattern classes in the dynamic case ($\mu_\eff>0$) can be attributed to different types of evolving pathogens (e.g., influenza, HIV, etc), and the patterns within a class as different variants of a given pathogen.   In our model, we will use the mutation rate as an order parameter  to characterize the different phases of  memory strategies in  biological systems.

\subsection{Optimal learning rate for evolving patterns}
In the classical Hopfield model ($\mu_\eff=0$) the learning rate $\lambda$ is set to very small values for the network to efficiently learn the patterns~\cite{workingMem87}. For evolving patterns, the learning rate should be tuned so the network can efficiently update the memory retained from the prior learning steps. At each encounter, the overlap $q^\alpha (t;\lambda)= \braket{ \sigma^\alpha_\att(t;\lambda)|\sigma^\alpha(t)}$ between a  pattern $\sigma^\alpha(t)$  and the corresponding attractor for the associated energy minimum  $\sigma_\attr^\alpha(t;\lambda)$ determines the accuracy of pattern recognition; the parameter $\lambda$ explicitly indicates the dependency of the network's energy landscape on the learning rate. 
We declare pattern recognition  as  successful if the accuracy of reconstruction (overlap) is larger than a set threshold $q^\alpha (t) \geq 0.8$, but  our results are insensitive to the  exact value of  this threshold (Appendix A and Fig.~\ref{fig:Dist_dist_paths}). We define a network's performance as the  asymptotic accuracy of its associative memory averaged over the ensemble of pattern classes (Fig.~\ref{Fig2}A),
\EQ
 \Q (\lambda) \equiv \E\left[q^\alpha(t;\lambda)\right] \simeq \lim_{T\to \infty} \frac{1}{T}\sum_{t=0}^T\frac{1}{N}\sum_{\alpha=1}^N\braket{ \sigma_\attr^\alpha(t)|\sigma^\alpha(t)}\label{eq.accuracy}
\EE
The expectation $\E[\cdot]$ is an empirical average over the ensemble of presented pattern classes  over time, which in the stationary state approaches the asymptotic average of the argument. The optimal learning rate is determined by maximizing the network's performance,  $ \lambda^* =  \argmax_\lambda {\Q}(\lambda)  $.

The optimal learning rate  increases with growing mutation rate so that a network can follow the evolving patterns (Fig.~\ref{Fig2}B). Although it is difficult to analytically calculate the  optimal learning rate, we can use an approximate approach and find the learning rate that minimizes the expected energy of the patterns $\av{E_{\lambda,\rho}(J,\sigma)}$, assuming that patterns are shown to the network at a fixed order (Appendix B). In this case, the expected energy is given by
\begin{equation}
\label{eq:expected_energy_order}
\av{E_{\lambda,\rho}(J,\sigma)} = \frac{L-1}{2} \times \frac{\lambda}{1 - \lambda}\times\frac{1}{\rho^{-2N}  (1 - \lambda)^{-N}-1 \,},
\end{equation}
 where {$\rho^N \equiv (1-2 \mu)^N \approx 1-2\mu_{\eff}$ } is the upper bound for the overlap between a pattern and its evolved form, when separated by the other $N-1$ patterns that are shown in between. The expected energy grows slowly with increasing mutation rate (i.e., with decreasing overlap $q$), and the approximation in eq.~\ref{eq:expected_energy_order} agrees very well with the numerical estimates for the scenario where pattens are shown in a random order (Fig.~\ref{Fig2}C). In the regime where memory can still be associated with the evolved patterns ($\mu_{\eff} \ll 0.5$), minimization of the expected energy (eq.~\ref{eq:expected_energy_order}) results in an optimal learning rate, 
\EQ \lambda^*(\mu) = \sqrt{8 \mu/(N-1)},\label{eq:lambda}\EE
that scales with the square root of the mutation rate. Notably, this theoretical approximation agrees well with the numerical estimates (Fig.~\ref{Fig2}B).

\subsection{Reduced accuracy of distributed associative memory against evolving patterns}
Despite using an optimized learning rate, a network's accuracy in pattern retrieval $\Q(\lambda)$  decays much faster than the na\"ive  expectation solely based on the  evolutionary divergence of patterns between two encounters with a given class (i.e., $ \Q_0=(1-2\mu)^N \approx  1-2\mu_{\eff}$); see Fig.~\ref{Fig2}A. There are two reasons for this reduced accuracy: (i) the lag in the network's memory against evolving patterns, and (ii) misclassification of presented patterns. 

The memory attractors associated with  a given pattern class can lag behind and only reflect the older patterns  presented prior to the most recent encounter of the network with the specified class. We characterize this lag $g_{\lag}$ by identifying a previous version of the pattern that has the maximum overlap with the network's energy landscape at a given time~$t$: 
$g_{\lag} = \argmax_{g \geq 0} \E\left[ \braket{\sigma(t-g \,N)|J(t)|\sigma(t-g\, N)}\right]$ (Appendix B.2). $g_{\lag}$ measures time  in units of $N$ (i.e., the effective separation time of pattern of the same class). An increase in the optimal  learning rate reduces  the time lag and enables the network to follow the evolving patterns more closely (Fig.~\ref{fig:Simresults_overleps_high}).  The accuracy of the  memory subject to such a lag decays as $\Q_\lag=  \rho^{g_\lag N} \approx1- 2 g_\lag  \mu_{\eff}$, which is faster than the na\"ive expectation (i.e., $1-2\mu_\eff$); see Fig.~\ref{Fig2}A. This memory lag explains the loss of performance for patterns that are still reconstructed by the network's memory attractors (i.e., those with $q^\alpha>0.8$; Fig.~\ref{fig:Simresults_overleps_high}A). However, the overall performance of the network  $\Q(\lambda)$ remains lower than the expectation obtained by taking into account this time lag (Fig.~\ref{Fig2}A)---a discrepancy that leads us to the second root of reduction in accuracy, i.e., pattern misclassification.

As the learning rate increases, the structure of the network's energy landscape changes. In particular, we see that with large learning rates a few narrow paths emerge between the memory attractors of networks (Fig.~\ref{fig:Fig1}C). As a result, the equilibration process for pattern retrieval can drive a presented pattern through the connecting paths towards a wrong memory attractor (i.e., one with a small overlap $\braket{\sigma_\attr|\sigma}$), which leads to pattern misclassification (Fig.~\ref{fig:Performance_no_evo}A and Fig.~\ref{fig:Dist_dist_paths}A,C). These paths  are narrow as there are only a few possible spin-flips (mutations) that can drive a pattern from one valley to another during equilibration (Fig.~\ref{fig:Dist_dist_paths}B,D and Fig.~\ref{fig:Paths_partisipation}A,C).   In other words, a large learning rate carves narrow mountain passes in the network's energy landscape (Fig.~\ref{fig:Fig1}C), resulting in  a growing fraction of  patterns to be misclassified.  Interestingly, pattern misclassification occurs even in the absence of mutations for networks with an increased learning rate (Fig.~\ref{fig:Performance_no_evo}A). This suggests that mutations only indirectly contribute to the misclassification of memory, as they necessitate a larger learning rate for the networks to optimally operate, which in turn results in the emergence of mountain passes in the energy landscape.

To understand  the memory misclassification, particularly for patterns with  moderately low (i.e.,  non-random) energy (Fig.~\ref{Fig2}C),  we use spectral decomposition to characterize the relative positioning of  patterns in the energy landscape (Appendix C). The vector representing each pattern $\ket{\sigma}$ can be expressed in terms of the network's eigenvectors $\{\Phi^i\}$, $\ket{\sigma} =\sum_i m_i \ket{\Phi^i}$, where the overlap $m_i \equiv \braket{\Phi^i|\sigma}$ is the $i^{th}$ component of the pattern in the network's coordinate system. During equilibration, we flip individual spins in a pattern and accept the flips based on their contribution to the recognition energy. We can view these spin-flips as rotations of the pattern in the space spanned by the eigenvectors  of the network. Stability of a pattern depends on whether these rotations could carry the pattern from its original subspace over to an alternative region associated with a different energy minimum.  

There are two key factors that modulate the stability of a pattern in a network. The  dimensionality of the subspace in which a pattern resides, i.e.,  support of a pattern by the network's eigenvectors, is one of the key determining factors for pattern stability. We quantify the support of a pattern $\sigma$ using the participation ratio~$\pi (\sigma) = (\sum_i m_i^2)^2/\sum_i m_i^4$~\cite{Bouchaud:1997pr,Derrida:1997rd}   that counts the number of eigenvectors that substantially overlap with the pattern. A small support~$\pi(\sigma)\approx 1$ indicates that the pattern is spanned by only a few eigenvectors and is restricted to a small sub-space, whereas a large support indicates that the pattern is orthogonal to only a few eigenvectors. As the learning rate increases, patterns  lie in lower dimensional sub-spaces supported by only a few eigenvectors (Fig.~\ref{fig:Paths_partisipation}B,D). This effect is exacerbated by the fact the energy gap between the eigenstates of the network also broaden with increasing learning rate (Fig.~\ref{fig:eigenvalues}). The combination of a smaller support for  patterns and a larger energy gap in networks with increased learning rate leads to the destabilization of patterns by enabling the spin-flips during equilibration to drive a  pattern from one subspace to another, through the mountain passes carved within the landscape; see Appendix C and Fig.~\ref{fig:Scatter_current_next} for the exact analytical criteria for pattern stability.

\subsection{Compartmentalized learning and memory storage}

Hopfield-like networks can store  accurate associative memory  for static patterns. However, these networks fail  to perform and store retrievable associative memory for evolving patterns (e.g. pathogens), even when learning is done at an optimal rate~(Fig.~\ref{Fig2}). To overcome this difficulty, we propose to store memory in compartmentalized networks, with  $C$ sub-networks of size $L_\C$ (i.e., the number of  nodes in a sub-network). Each compartment (sub-network) can store a few of the total of $N$ pattern classes without an interference from  the other compartments (Fig.~\ref{fig:Fig1}D). 

Recognition of a pattern $\sigma$ in a compartmentalized network involves a two step process (Fig.~\ref{fig:Fig1}D): First, we choose a  sub-network $J^i$ associated with compartment $i$ with a probability $P_i =\exp[-\beta_\rS E(J^i,\sigma)]/\mathcal{N}$, where $\beta_\rS$ is the inverse temperature for  this decision and~$\mathcal{N}$ is the normalizing factor. Once the compartment is chosen, we  follow the recipe for pattern retrieval  in the energy  landscape of the associated sub-network, whereby a pattern equilibrates into a  memory attractor.

On average, each compartment  stores a memory for $N_\C=N/C$ pattern classes.  To keep the networks with different number of compartments $C$ comparable, we scale the  size of each compartment $L_\C$ to keep $C\times L_\C =\text{constant}$, which keeps the (Hopfield) capacity of the network $\alpha =  {N_\C}/{L_\C}$ invariant under compartmentalization. Moreover, the mutation rate experienced by each sub-network  scales with the number of compartments $\mu_\C= C \mu$, which keeps the  effective mutation rate $\mu_\eff = N_\C\mu_\C$ invariant  under compartmentalization. As a result, the  optimal learning rate (eq.~\ref{eq:lambda}) scales with the  number of compartments~$C$ as, $\lambda_\C^* = \sqrt{8 \mu_\C/(N_\C-1})\approx C\lambda_1^*$. However, since updates are restricted to sub-networks of size $L_\C$ at a time, the expected amount of updates within a network $L_\C\lambda_\C$ remains invariant under compartmentalization. Lastly, since the change in energy by a single spin-flip scales as $\Delta E \sim 1/L_\C$, we introduce the scaled  Hopfield temperature $\beta_{\rH_\C} \equiv C\beta_\rH $ to make the equilibration process comparable across networks with different number of compartments.  No such scaling is necessary  for $\beta_\rS$.

By restricting the networks to satisfy the aforementioned scaling relationships, we are left with two independent variables, i.e., (i) the number of compartments $C$, and (ii) the learning rate $\lambda_\C$, which define a memory strategy $\{C,\lambda_\C\}$. A memory strategy  can be then optimized to achieve a maximum accuracy for retrieving an associative memory for evolving  patterns with a given effective mutation rate $\mu_\eff$.

\subsection{Phases of learning and memory production}
Pattern retrieval can be stochastic due to the noise in choosing the right compartment from the $C$ sub-networks (tuned by the inverse temperature $\beta_\rS$), or the noise in equilibrating into the right memory attractor in the energy landscape of the chosen sub-network (tuned by the Hopfield inverse temperature $\beta_{\rH_\C}$). We use mutual information to quantify the accuracy of patten-compartment association, where larger values indicate a more accurate association; see Appendix A and Fig.~\ref{Fig4}. The optimal performance $\Q^*$ determines the overall accuracy of memory retrieval, which depends on both finding the right compartment and equilibrating into the correct memory attractor.  The amplitudes of intra- versus inter- compartment stochasticity determine the optimal strategy $\{C^*,\lambda^*_\C\}$ used for learning and retrieval of patterns with a specified mutation rate. Varying the corresponding inverse temperatures ($\beta_{\rH_\C},\beta_\rS$) results in three distinct phases of optimal memory storage.\\

\begin{figure}[t!]
\includegraphics[width=\columnwidth]{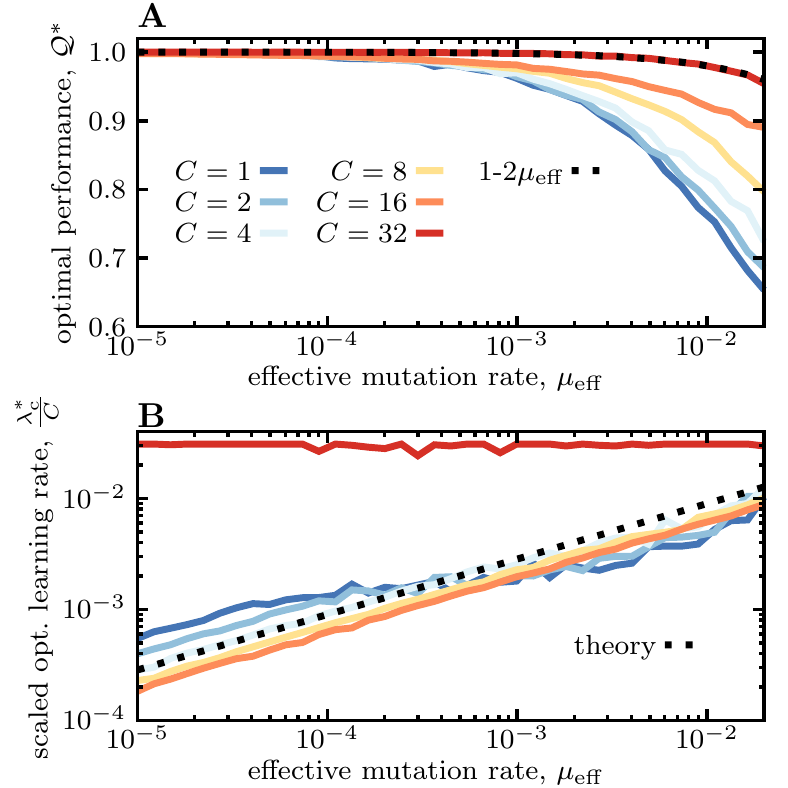}
\caption{
\label{Fig3}
{\bf Compartmentalized memory storage.}
{\bf (A)} The optimal performance is shown as a function of the  effective mutation rate (similar to Fig.~\ref{Fig2}A) for networks with different number of  compartments $C$ (colors), ranging from a network with distributed memory $C=1$ (blue) to a 1-to-1 compartmentalized network $C=N$ (red). {\bf (B)} The optimal (scaled) learning rate  $\lambda_\C/C$  is shown as a function of the effective mutation rate for networks with different number of compartments (colors according to (A)). Full lines show the numerical estimates and the dashed line is from the analytical approximation, $\lambda_\C^*=\sqrt{8 \mu_\C/(N_\C-1})\approx C\lambda_1^*$.  The scaled learning rates collapse on the analytical approximation for all networks except for the 1-to-1 compartmentalized network (red), where the maximal learning rate $\lambda \approx 1$ is used and each compartment is fully updated upon an encounter with a new version of a pattern. The number of presented patterns is set to $N=32$. We keep $L\times C = \text{const.}$, with $L=800$ used for the network with $C=1$.
}
\end{figure}

\begin{figure*}[t!]
\includegraphics[width=\columnwidth]{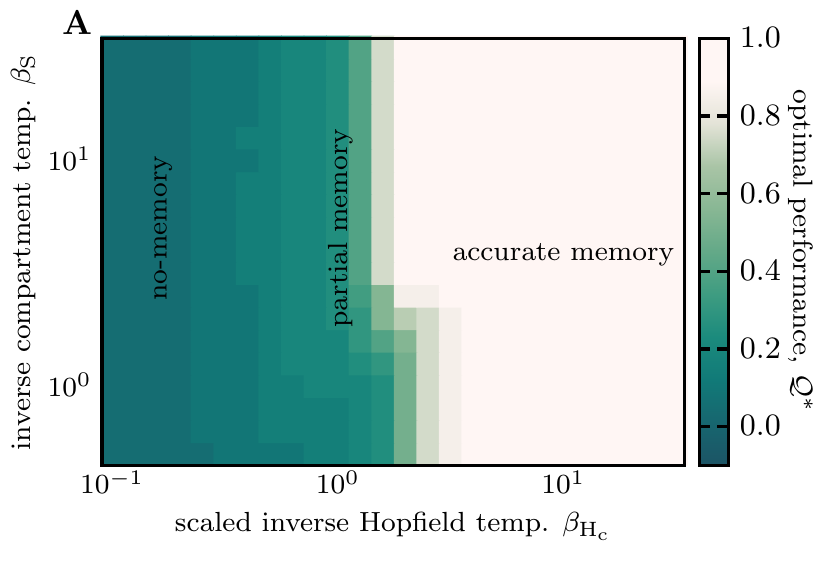}
\includegraphics[width=\columnwidth]{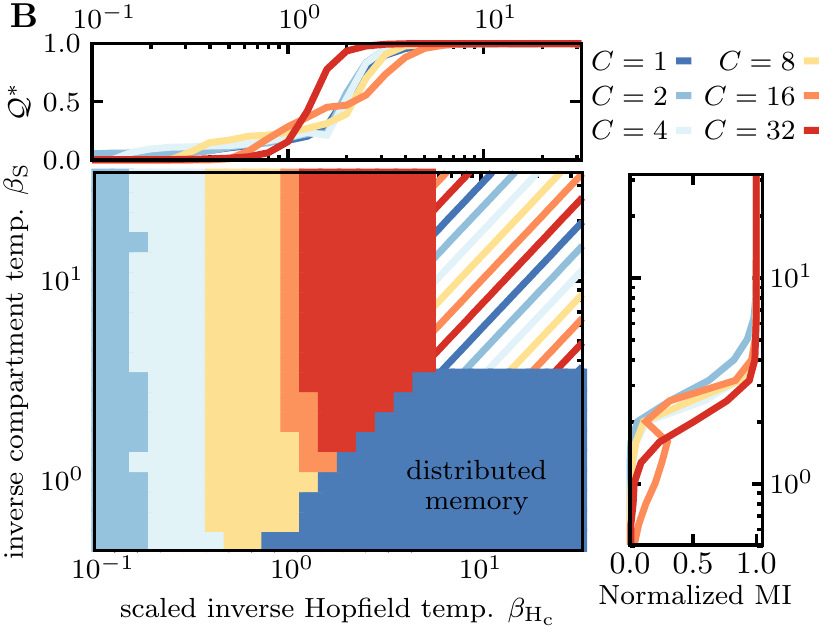}
\includegraphics[width=\columnwidth]{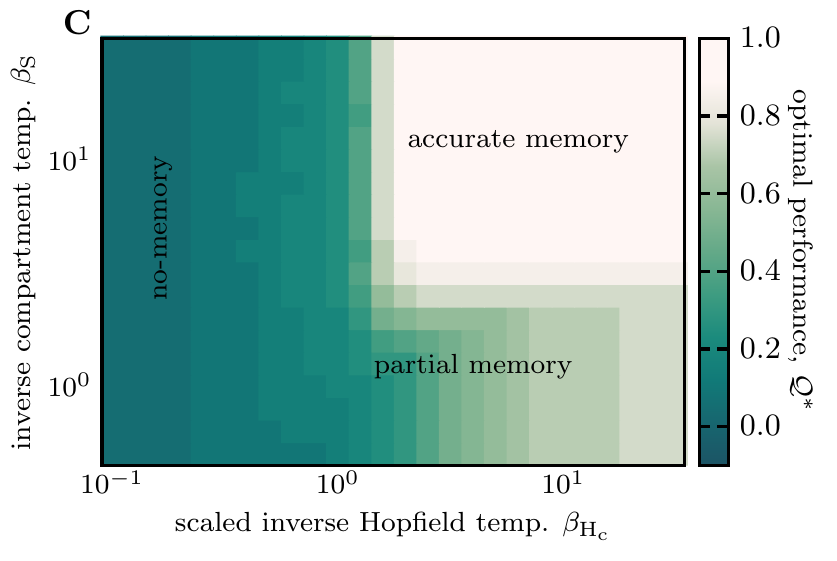}
\includegraphics[width=\columnwidth]{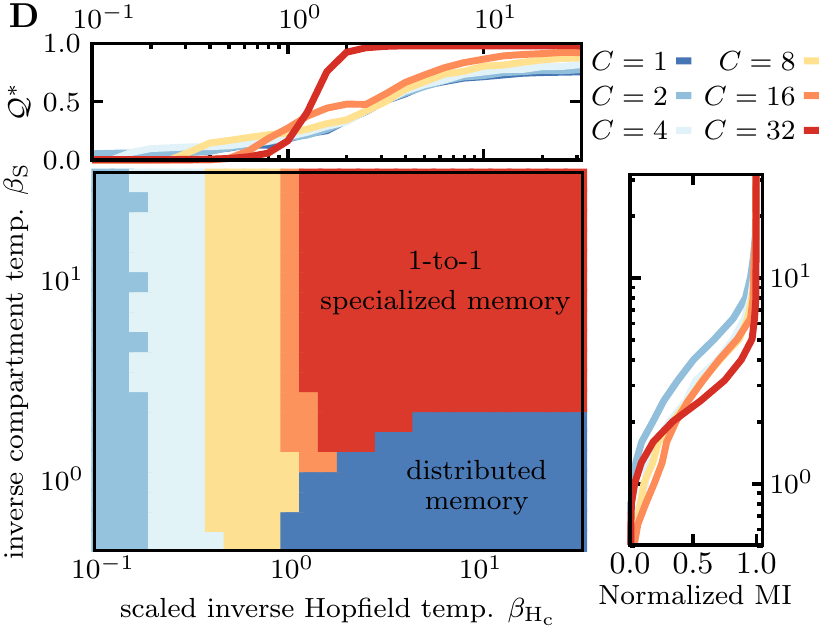}
\caption{\label{Fig4}
{\bf  Phases of learning and memory production.} 
Different optimal memory strategies are shown. {\bf (A)} The heatmap shows the optimal memory performance $\Q^*$ as a function of the (scaled) Hopfield inverse temperature $\beta_{\rH_\C} = \beta_\rH\cdot C$ and the inverse temperature associated with compartmentalization $\beta_\rS$ for networks learning and retrieving a memory of static patterns ($\mu =0$); colors indicated in the color bar. The optimal  performance is achieved by using the optimal strategy (i.e., learning rate $\lambda_\C^*$  and the number of compartments $\C^*$) for networks at each value of $\beta_{\rH_\C} $ and $\beta_\rS$. 
The three phases of accurate, partial, and  no-memory are indicated.  {\bf (B)} The heatmap shows the memory strategies for the optimal number of compartments (colors as in the legend) corresponding to the memory performance shown in (A). We limit the optimization to the possible number of compartments indicated in the legend to keep $N/C$ an integer. The dashed region corresponds to the case where all  strategies perform equally well. Regions of distributed memory ($C=1$) and the 1-to-1 specialized memory ($C=N$) are indicated. The top panel shows the optimal performance $\Q^*$ of different strategies   as a function of the Hopfield inverse temperature $\beta_{\rH_\C}$. The right panel shows the mutual information $\text{MI}(\Sigma,\mathcal{C})$ between the set of pattern classes $\Sigma \equiv \{\sigma^\alpha \}$ and the set of all compartments  $\mathcal{C}$  normalized by the entropy of the compartments $\rH(\mathcal{C})$  as a function of the inverse temperature $\beta_\rS$; see Appendix A.3. This normalized mutual information quantifies the ability of the system to  assign a pattern  to the correct compartment. {\bf (C-D)} Similar to (A-B) but for evolving patterns with the effective mutation rate $\mu_\eff= 0.01$. The number of presented patterns  is set to  $N=32$ (all panels). Similar to Fig.~\ref{Fig3} we keep $L\times C = \text{const.}$, with $L=800$ used for  networks with $C=1$. }
\end{figure*}

 \noindent{\bf \small i.~Small intra- and inter-compartment noise ($\beta_{\rH_\C} \gg 1$, $\beta_\rS \gg1$).}
In this regime, neither the compartment choice nor the pattern retrieval within a compartment are subject to strong noise. As a result,  networks are functional with working memory and the optimal strategies can achieve the highest overall performance. For small mutation rates, we see that all networks perform equally well and can achieve almost perfect performance, irrespective of their number of compartments (Figs.~\ref{Fig3}A,~\ref{Fig4}A,B). As the mutation rate increases, networks with a larger number of compartments show a more favorable performance, and the 1-to-1 specialized network, in which each pattern is stored in a separate compartment (i.e., $N=C$), reaches the optimal performance $1-2\mu_\eff$ (Figs.~\ref{Fig3}A,~\ref{Fig4}C,D). As predicted by the theory, the optimal learning rate for compartmentalized networks scales with the mutation rate as  $\lambda^*_\C\sim\mu_\C^{1/2}$, except for the 1-to-1 network in which $\lambda^*_\C\to 1$ and sub-networks are steadily updated upon an encounter with a pattern (Fig.~\ref{Fig3}B). This rapid update is expected since there is no interference between the stored memories in the 1-to-1 network, and a steady update can keep the stored memory in each sub-network close to its associated pattern class without disrupting the other energy minima. \\

\noindent {\bf \small ii.~Small intra- and large inter-compartment noise ($\beta_{\rH_\C} \gg 1$, $\beta_\rS \ll 1$).}
In this regime there is low noise for equilibration within a compartment but a high level of noise in choosing the right compartment.  The optimal strategy in this regime is to store  patterns in a single network with a distributed memory, since identifying the correct compartment is difficult due to noise (Fig.~\ref{Fig4}B,D). For static patterns this strategy corresponds to the classical Hopfield model with a high accuracy (Figs.~\ref{Fig2}A,~\ref{Fig4}A,B). On the other hand, for evolving patterns  this strategy results in a partial memory (Fig.~\ref{Fig4}C,D)  due to  the reduced accuracy of the distributed associative memory, as shown in Fig.~\ref{Fig2}A. Interestingly, the transition between the optimal strategy with highly specific (compartmentalized) memory for evolving patterns in the first regime and the generalized (distributed) memory in this regime is very sharp (Fig.~\ref{Fig4}D). This sharp transition suggests that depending on the noise in choosing the compartments $\beta_\rS$, an optimal  strategy either stores  memory in  a 1-to-1 specialized fashion ($C=N$) or in a distributed generalized fashion ($C=1$), but no intermediate solution (i.e., quasi-specialized memory with $1<C<N$) is desirable. \\

\noindent {\bf \small iii.~Large intra-compartment noise ($\beta_{\rH_\C} < 1$).}
In this regime there is a high level of noise in equilibration within a network and memory cannot be reliably retrieved (Fig.~\ref{Fig4}A,C), regardless of the compartmentalization temperature $\beta_\rS$. However, the impact of the equilibration noise $\beta_{\rH_\C}$ on the accuracy of memory retrieval depends on the degree of compartmentalization. For the 1-to-1 specialized network ($C=N$), the transition between the high and the low accuracy is {smooth} and occurs at $\beta_{\rH_\C}=1$, below which no memory attractor can be found. As we increase the  equilibration noise (decrease $\beta_{\rH_\C}$), the networks with distributed memory ($C<N$) show two-step transitions, with a plateau in the range of $1/N_\C\lesssim\beta_{\rH_\C}\lesssim 1$. Similar to the 1-to-1 network, the first transition at $\beta_{\rH_\C}\approx 1$ results in the reduced accuracy of the networks' memory retrieval. At this transition point, the networks' learning rate $\lambda_\C$ approaches its maximum value $1$ (Fig.~\ref{fig:Phasetransition_Beta_H_SI}), which implies that the memory is stored (and steadily updated) for only $C <N$ patterns (i.e., one pattern per sub-network). Due to the invariance of the networks' mean energy under compartmentalization, the depth of the energy minima associated with the stored memory in each sub-network scales as $N/C$, resulting in deeper and more stable energy minima in networks with smaller number of compartments $C$. Therefore, as the noise increases (i.e., $\beta_{\rH_\C}$ decreases), we observe a gradient in transition from partial retrieval to a no-memory  state at $\beta_H\approx 1/N_\C$, with the most compartmentalized network (larger $C$) transitioning the fastest, reflecting the shallowness of its energy minima. 

Taken together, the optimal strategy leading to working memory  depends on whether a network  is trained to learn and retrieve dynamic (evolving) patterns or static patterns. Specifically, we see that the 1-to-1 specialized network is the unique optimal solution for storing working memory for evolving patterns, whereas the distributed generalized memory (i.e., the classical Hopfield network) performs equally well in learning and retrieval of  memory for static patterns. The contrast between these memory strategies can shed light on the distinct molecular mechanisms utilized by different biological systems to store memory.

\section{Discussion}
Storing and retrieving memory from prior molecular interactions is an efficient scheme to sense and respond to external stimuli. Here, we introduced a flexible energy-based network model that can adopt different memory strategies, including distributed memory, similar to the classical Hopfield network, or compartmentalized memory. The  learning rate and the number of compartments in a network define a memory strategy, and we probed the efficacy of different strategies  for static and dynamic patterns.  We found that Hopfield-like networks with distributed memory are highly accurate in storing associative memory for static patterns. However,  these networks fail to reliably store retrievable associative memory for evolving patterns, even when learning is done at an optimal rate. 

To achieve a high accuracy, we showed that a retrievable memory for evolving patterns should be compartmentalized, where each pattern class is stored in a separate sub-network. In addition, we found a sharp transition between the different phases of working memory (i.e., compartmentalized and distributed memory), suggesting that  intermediate solutions (i.e., quasi-specialized memory) are sub-optimal against evolving patterns. 

The contrast between these memory strategies is reflective of the distinct molecular mechanisms used for memory storage in the adaptive immune system and in the olfactory cortex.
 In particular, the memory of odor complexes, which can be assumed as static, is stored in a distributed fashion in the olfactory cortex~\cite{Haberly:1989jo,Brennan:1990bp,Granger:1991hu,Haberly:2001gx,Wilson:2004do,Lansner:2009hb}. On the other hand, the  adaptive immune system, which encounters evolving pathogens, allocates distinct immune cells (i.e., compartments) to  store a memory for different types of pathogens (e.g. different variants of  influenza or HIV)---a strategy that resembles that of the  1-to-1 specialized networks~\cite{Janeway:2001te,Mayer:2015ce,Shinnakasu:2016ei,Shinnakasu:2017ct,Mayer:2019is,Schnaack:2020vb,Viant:2020jt}. Our results suggest that pathogenic evolution may be one of the reasons for the immune system to encode a specialized memory, as opposed to the distributed memory used in the olfactory system.

The increase in the optimal learning rate  in anticipation of patterns' evolution significantly changes the structure of the energy landscape for associative memory. In particular, we found the emergence of narrow connectors (mountain passes) between the memory attractors of a network, which destabilize the equilibration process and significantly reduce the accuracy of memory retrieval. Indeed, tuning the learning rate as a hyper-parameter is one of the challenges of current machine learning algorithms with deep neural networks (DNNs)~\cite{Goodfellow-et-al-2016,Mehta:2019pr}. The goal is to navigate the tradeoff between the speed (i.e., rate of convergence) and accuracy without overshooting during optimization. It will be interesting to see how the insights developed in this work can inform rational approaches to  choose an optimal learning rate in optimization tasks  with DNNs. 

Machine learning algorithms with DNNs~\cite{Goodfellow-et-al-2016} and modified  Hopfield networks~\cite{parga:1986jp,Virasoro:1986gk,gutfreund:1988pr,tsodyks:1990mp} are able to accurately classify hierarchically correlated patterns, where different  objects can be organized  into an ultrametric tree based on some specified relations of similarity. For example, faces of  cats and dogs have the oval shape in common but they branch out in the ultrametric tree according to the organism-specific features, e.g., whiskers in a cat, and the cat branch can then further diversify based on the breed-specific features. A classification algorithm can use these hierarchical relations to find features  common among members of a given  sub-type (cats) that can distinguish them from another sub-type (dogs).  Although evolving patterns within each class are correlated, the random evolutionary dynamics of these patterns does not build a hierarchical structure where a pattern class branches in two sub-classes that share a common ancestral root.  Therefore, the optimal memory strategies found here for evolving patterns are distinct from those of the hierarchically correlated patterns. It will be interesting to see how  our approaches can  be implemented in DNNs to classify dynamic and evolving patterns. 

\section*{Acknowledgement}
This work has been supported by the DFG grant (SFB1310) for Predictability in Evolution and the MPRG funding through the Max Planck Society. O.H.S also acknowledges funding from Georg-August University School of Science (GAUSS) and the Fulbright foundation. 

\clearpage{}
\newpage{}

\onecolumngrid

%\noindent{\Large \bf Materials and Methods}\\\\

\renewcommand{\theequation}{S\arabic{equation}} 

\appendix

  \setcounter{figure}{0}
\renewcommand{\thefigure}{S\arabic{figure}}
\renewcommand{\theequation}{S\arabic{equation}} 
\setcounter{section}{0}

\section{Computational procedures}
\renewcommand{\thesubsection}{A\arabic{subsection}}
\subsection{Initialization of the  network}
A network $J$ (with elements $J_{ij}$) is presented with $N$ random (orthogonal) patterns $\ket{\sigma^\alpha}$ (with $\alpha = 1,\dots N$), with entries $\sigma^{\alpha}_i \in \{-1,1\}$, reflecting the $N$ pattern classes. 
For a network with  $ C$ compartments (with $1\leq C\leq N$), we initialize each sub-network $J^s$  at time $t_0$ as $J^s_{i,j}(t_0) =  \frac{1}{N/C} \sum_{\alpha  \in \mathcal{A}_s} \sigma^{\alpha}_i  \sigma^{\alpha}_j$ and $J^s_{ii}(t_0)=0$; here,  $\mathcal{A}_s$ is a set of $N/C$ randomly chosen (without replacement) patterns initially assigned to the compartment (sub-network) $s$.  We then let  the network undergo an initial learning process.  At each step an arbitrary  pattern $\sigma^{\nu}$ is  presented to the network and a sub-network $J^s$ is chosen for an update with a  probability \begin{align}
 \label{eq:choose_compartmemt_SI}
  P_s = \frac{\exp\left[{-\beta_\rS E\left(J^s(t),\sigma^\nu(t)\right)}\right]}{\sum_{r=1}^{C}\exp\left[{-\beta_\rS   E\left(J^s(t),\sigma^\nu(t)\right)}\right]},
 \end{align}
where the energy is defined as, \EQ 
\label{eq:energy_def_SI}
E\left(J^s(t),\sigma^\nu(t)\right) = \frac{-1}{2L}\sum_{i,j} J^s_{i,j}(t)\sigma^\nu_i(t) \sigma^\nu_j(t) \equiv \frac{-1}{2}\braket{\sigma^\nu(t)|J^s(t)|\sigma^\nu(t) }\EE
 and $\beta_\rS  $ is the inverse temperature associated with choosing the right compartment. We then update the selected  sub-network $J^s$, using the Hebbian update rule, 
\begin{equation}
J^s_{i,j}(t+1) = 
\begin{cases}
(1-\lambda)\,J^s_{i,j} (t) +\lambda \,\sigma^{\nu}_i\sigma^{\nu}_j,&\text{if }i\neq j;\\
0,&\text{otherwise.}
\end{cases}
\label{eq:learningrule_SI}
\end{equation}
For dynamic patterns, the presented patterns undergo evolution with mutation rate $\mu$, which reflects the average number of spin flips in a given pattern per network's update event (Fig.~\ref{fig:Fig1}).  

Our goal is to study the memory retrieval problem in a network that has reached its steady state. The state of a network  $J(t_n)$ at the time step $n$ can be traced back to the initial state $J(t_0)$ as,
\EQ
J(t_n) = (1-\lambda)^n J(t_0) + \lambda \sum_{i=1}^n   (1-\lambda)^{n-i} \ket{\sigma(t_{i})}\bra{\sigma(t_{i})}
\label{eq.Jn0_SI}\EE

The contribution of the initial state  $J(t_0)$ to the state of the network at time  $t_n$ decays as $(1-\lambda)^n$ (eq.~\ref{eq.Jn0_SI}). Therefore, we choose the number of steps to reach the steady state as  $n_{\rm stat.} = \max \left[10 N,  2C\,\operatorname{ceil} \left(  \frac{\log10^{-5}}{\log(1-\lambda)} \right) \right]$. {This criteria ensures that $(1-\lambda)^{n_{\rm stat.} }\leq 10^5$ and the memory of the initial state $J(t_0)$ is removed from the network $J(t)$.}
We will then use this updated network to collect the statistics for memory retrieval. To report a specific quantity from the network (e.g., the energy), we pool the $n_{\rm stat.}$ samples collected from each of the 50 realizations.

 \subsection{Pattern retrieval from associative memory}
Once the trained network approaches a stationary state, we collect the  statistics of the stored memory.   \\
 
To find a memory attractor $\sigma^\alpha_{\text{att}}$ for a given pattern $\sigma^\alpha$ we  use a Metropolis algorithm in the energy landscape $E(J^s,\sigma^\alpha)$  (eq.~\ref{eq:energy_def_SI}). To do so, we make spin-flips in a presented pattern $\sigma^\alpha \rightarrow \tilde{\sigma}^\alpha$ and accept a spin-flip   with probability
\begin{align}
\label{eq:gradiant_descent_Si}
P(\sigma^\alpha \rightarrow \tilde{\sigma}^\alpha) = \min \left( 1 , e^{- \beta_\rH \Delta E} \right),
\end{align}
where $\Delta E = E(J^s,\tilde{\sigma}^\alpha) -E(J^s,\sigma^\alpha) $ and $\beta_\rH$ is the inverse (Hopfield) temperature for pattern retrieval in the network (see Fig.~\ref{fig:Fig1}).  We repeat this step for $2 \times 10^{6}$ steps, which is sufficient to find a minimum of the landscape (see Fig.~\ref{fig:Dist_dist_paths}). 

For systems with more than one compartment $C$, we first choose a compartment according to eq.~\ref{eq:choose_compartmemt_SI}, and then perform the Metropolis algorithm within the associated compartment.

After finding the energy minima, we  update the systems for $n'_{\rm stat.} = \max [2\cdot 10^3,n_{\rm stat.}]$ steps. At each step we present patterns as described above and collect the statistics of the recognition energy $E(J^s(t),\sigma^\alpha(t))$ between a presented pattern $\sigma^\alpha$ and the memory compartment $J^s(t)$, assigned according to eq.~\ref{eq:choose_compartmemt_SI}. These measurements are used to describe the energy statistics {(Figs.~\ref{Fig2},\ref{fig:Performance_no_evo})} of the patterns and the accuracy of pattern-compartment association {(Fig.~\ref{Fig4}B,D)}. After the $n'_{\rm stat.} $ steps, we again use the  Metropolis algorithm to find the memory attractors associated with the presented patterns. We repeat this analysis for $50$ independent realizations of the initializing pattern classes $\{\sigma^\alpha(t_0)\}$, for each set of parameters $\{ L,N,C,\lambda,\mu,\beta_\rS  , \beta_\rH \}$. 

When calculating the mean performance $\Q$ of a strategy (see Figs.~\ref{Fig2},\ref{Fig3},\ref{Fig4},\ref{fig:Phasetransition_Beta_H_SI}), we set the overlap between attractor and pattern $q^\alpha = |\braket{\sigma^{\alpha}_\att|\sigma^\alpha }|$ equal to zero when patterns are not recognized ($q^\alpha<0.8$). As a result, systems can only achieve a non-zero performance if they recognize some of the patterns. This choice eliminates the  finite size effect of a random overlap $\sim 1 / \sqrt{L}$ between an attractor and a pattern (see Fig.~\ref{fig:Dist_dist_paths}). This correction is especially important when comparing systems with different sub-network sizes ($L_\C \equiv L/C$) in the $\beta_\rH <1$  regime  (Figs.~\ref{Fig4},\ref{fig:Phasetransition_Beta_H_SI}), where random overlaps for small $L_\C$ could otherwise result in a larger mean performance compared to larger systems that correctly reconstruct a fraction of the patterns.

\subsection{Accuracy of pattern-compartment association}
We use the mutual information $\MI(\Sigma,\mathcal{C})$ between the set of pattern classes $\Sigma \equiv \{\sigma^\alpha \}$ and the set of all compartments  $\mathcal{C}$ to  quantify the accuracy in associating a presented pattern with the correct  compartment, 
\EQ
\begin{split}
\label{SI:MI}\MI(\Sigma,\mathcal{C}) &=   H(\mathcal{C}) - H(\mathcal{C}|\Sigma)\\
 &= - \sum_{c\in \mathcal{C}}  P(c) \log P(c) - \left[- \sum_{\sigma^\alpha\in\Sigma} P(\sigma^\alpha)\sum_{c\in\mathcal{C}}  P(c|\sigma^\alpha) \log P(c|\sigma^\alpha) \right].
\end{split}
\EE
Here $H(\mathcal{C})$ and $H(\mathcal{C}|\Sigma) $  are the entropy of the  compartments, and the conditional entropy of the compartments given the presented patterns, respectively. If chosen randomly, the entropy associated with choosing a compartment is $H_\text{ random}(\mathcal{C}) = \log C$.  The mutual information (eq.~\ref{SI:MI})  measures the reduction in the entropy of compartments due to the association between the patterns and the compartments, measured by the conditional entropy $H(\mathcal{C}|\Sigma)$.  Figure 4B,D shows the mutual information $\MI(\Sigma,\mathcal{C})$ scaled by its upper bound  $H(\mathcal{C})$, in order to compare networks with a different number of compartments. 

\newpage{}
  \section{Estimating  energy and  optimal learning rate for working memory}
  \renewcommand{\thesubsection}{B\arabic{subsection}}
  \subsection{Approximate solution for optimal learning rate}
 The optimal learning rate is determined by maximizing the network's performance $\Q(\lambda)$ (eq.~\ref{eq.accuracy}) against evolving patterns with a specified mutation rate:  

 \EQ
 \lambda^* = \argmax_\lambda {\Q}(\lambda) 
 \label{eq.SI.lambda} 
 \EE  
 
We can  numerically estimate the optimal learning rate as defined eq.~\ref{eq.SI.lambda}; see Figs.~\ref{Fig2},\ref{Fig3}. However, the exact analytical evaluation of the optimal learning rate is difficult and we use  an approximate approach and find the learning rate that minimizes the expected energy of the patterns in the stationary state $\av{E_{{}_{\lambda,\rho}}(J,\sigma)}$, assuming that patterns are shown to the network at a fixed order. Here, the subscripts explicitly indicate the learning rate of the network $\lambda$, and the evolutionary overlap of the pattern $\rho$. To evaluate an analytical approximation for the energy, we first evaluate the state of the network $J(t)$ at time step $t$, given all the prior encounters of the networks with patterns shown at a fixed order. 
\begin{align} 
\label{eq:J-with_patterns_first_order} \frac{1}{L}\left(J (t) + \mathds{1}\right)  &= \lambda \sum_{j=1}^\infty (1-\lambda)^{(j-1)} \ket{\sigma(t-j)} \bra{\sigma(t-j)} \\
\label{eq:J-with_patterns_second order}&= \lambda \underbrace{\sum_{i=1}^\infty (1-\lambda)^{(i-1)N} \underbrace{\sum_{\alpha=1}^N  (1-\lambda)^{\alpha-1}\ket{\sigma^\alpha(t-\alpha - (i-1)N)} \bra{\sigma^\alpha(t-\alpha - (i-1)N)} }_{\text{sum over {\it N} pattern classes}}}_{\text{sum over time (generations, } i)} \\
&=\lambda  \underbrace{\sum_{\alpha=1}^N \underbrace{\sum_{i=0}^\infty (1-\lambda)^{(\alpha-1) + iN}\ket{\sigma^\alpha(t-\alpha - iN)} \bra{\sigma^\alpha(t-\alpha - iN)}}_{\text{sum over time}}  }_{\text{sum over patterns}}.
\end{align}
Here, we referred to  the (normalized) pattern vector  from the class $\alpha$ presented to the network at time step $t$ by $\ket{\sigma^\alpha(t)} \equiv \frac{1}{\sqrt{L} }\sigma^\alpha(t)$.   Without a loss of generality, we assumed that the  last  pattern presented to the network at time step $t-1$ is from the first pattern class $\ket{\sigma^1(t-1)}$, which enabled us to split the sum in eq.~\ref{eq:J-with_patterns_first_order} into two separate summations over pattern classes and $N$ time-steps generations (eq.~\ref{eq:J-with_patterns_second order}). Adding the identity matrix $\mathds{1} $ on the left-hand side of eq.~\ref{eq:J-with_patterns_first_order} assures that the diagonal elements vanish, as defined in eq.~\ref{eq:learningrule_SI}. 

The mean energy of the patterns, which  in our setup is  the energy of the pattern from the  $N^{ th}$ class at time $t$, follows
\begin{equation}
\begin{split}
\av{E_{{}_{\lambda,\rho}}(J,\sigma)}&=\av{-\frac{1}{2} \braket{\sigma^N(t)|J(t)|\sigma^N(t)}} \\
\label{eq:energypattern}
&= \av{-\frac{L-1}{2}  \lambda\sum_{\alpha=1}^N \sum_{i=0}^\infty  (1-\lambda)^{(\alpha-1) + iN} \braket{\sigma^N(t)|\sigma^\alpha(t-\alpha - iN)} \braket{\sigma^\alpha(t-\alpha - iN)|\sigma^N(t)}}.
\end{split}
\end{equation}

Since the pattern families are orthogonal to each other, we can express the  overlap between patterns at different times as $\braket{\sigma^\alpha(t_1)|\sigma^\nu(t_2)} = \delta_{\alpha,\nu} (1-2\mu)^{|t_2-t_1|} \equiv \delta_{\alpha,\nu} \rho^{|t_2-t_1|} $, and simplify the energy function in  eq.~\ref{eq:energypattern},
 \begin{equation}
 \begin{split}
\av{E_{{}_{\lambda,\rho}}(J,\sigma)}&=-\frac{L-1}{2}  \lambda\sum_{i=0}^\infty  (1-\lambda)^{(N-1) + iN} \rho^{2(N +iN)}\\
&= -\frac{L-1}{2}  \lambda    (1-\lambda)^{(N-1)}  \rho^{2N} \sum_{i=0}^\infty \left((1-\lambda)^{N}  \rho^{2N} \right)^i \\
\label{eq:energypattern_final}
&= -\frac{L-1}{2}\,  \lambda \,  \frac{(1-\lambda)^{(N-1)}  \rho^{2N}}{1 - (1-\lambda)^{N}  \rho^{2N} }.
\end{split}
\end{equation}

Since accurate pattern retrieval  depends on the  depth of the energy valley for the associative memory,  we will use the expected energy of the patterns as a proxy for the performance of the network. We can find the approximate optimal learning rate that minimized the expected energy by setting $\partial \av{E_{{}_{\lambda,\rho}}(J,\sigma)}/{\partial \lambda } =0$, which results in 
\begin{equation}
\begin{split}
\label{eq.SI.learning}(1-2 \mu)^{2N} = (1- \lambda^*)^{-N}  (1 -  N \lambda^*) &\Longrightarrow  1 - 4 N \mu + \mathcal{O}(\mu^2) = 1+ \frac{1}{2}(N-N^2) (\lambda^*)^2 + \mathcal{O}(\lambda^3);\\
&\Longrightarrow  \lambda^*(\mu) \simeq \sqrt{8 \mu/(N-1)}.
\end{split}
\end{equation}
where we used the fact that both the mutation rate $\mu$ and the learning rate $\lambda$ are small, and therefore, expanded eq.~\ref{eq.SI.learning} up to the leading orders in these parameters. 

In addition, eq.~\ref{eq.SI.learning}  establishes an upper bound for the learning rate:  $\lambda <\frac{1}{N}$. Therefore, our expansion in mutation rate (eq.~\ref{eq.SI.learning}) is only valid for $8\mu<\frac{1}{N}$, or equivalently for $\mu_{\text{eff}} = N \mu < 12.5 \%$;  the rates used in our analyses lie far below these upper bounds.  

\subsection{Lag of memory against evolving patterns}
The memory attractors associated with  a given pattern class can lag behind and only reflect the older patterns  presented prior to the most recent encounter of the network with the specified class.  As a result, the upper bound for the performance of a network $\Q_\lag=  \rho^{g_\lag N} \approx1- 2 g_\lag  \mu_{\eff}$ is determined by both the evolutionary divergence of patterns between two encounters {$\mu_\eff$} and number of generations  $g_\lag$ by which the stored memory  lags behind; we measure $g_\lag$ in units of generations;  one generation is defined as  the average time between a network's encounter with the same pattern class i.e., $N$. We characterize this lag $g_{\lag}$ by identifying the past pattern (at time $t- g_\lag N$) that has the maximum overlap with the network's energy landscape at given time~$t$: 
\EQ 
g_{\lag} =   \argmax_{g \geq 0}\E\left[ \braket{\sigma(t-g \,N)|J(t)|\sigma(t-g\, N)}\right] \equiv \argmin_{g \geq 0} \av {E_\lag(g)}
\EE
where we introduced the expected lagged energy $\av {E_\lag(g)}$. Here, the vector $\ket{\sigma(t)}$ refers to the pattern $\sigma$ presented to the network at time $t$, which can be from any of the pattern classes.
 Because of  the translational symmetry in time in the stationary state, the lagged energy  can also be expressed in terms of the overlap between a pattern at time $t$ and the network at a later time $t+ gN$. We  evaluate the lagged energy by substituting the expression for the network's state $J(t+ gN)$ from eq.~\ref{eq:J-with_patterns_first_order}, which entails,  
\begin{align}
\label{eq:energypattern_lag_1} \frac{2}{L-1} \av{E_{\text{lag}}(g)}&=  - \frac{1}{L-1} \av{\braket{\sigma(t)|J(t+g \,N)|\sigma(t)} }\\
\label{eq:energypattern_lag_2}  &= - \av{ \frac{1}{L-1}  (1-\lambda)^{Ng} \braket{\sigma(t)|J(t)|\sigma(t)} + \lambda \sum_{j=0}^{gN-1} (1-\lambda)^{gN-1-j}  \braket{\sigma(t)|\sigma(t+j)}^2 }\\ 
\label{eq:energypattern_lag_3}&=  \frac{2}{L-1} (1-\lambda)^{Ng} \av{E_{{}_{\lambda,\rho}}(J,\sigma)}   - \lambda     \sum_{i=0}^{g-1}\sum_{\alpha=0}^{N-1} (1-\lambda)^{gN-1-Ni-\alpha}  \braket{\sigma^{N}(t)|\sigma^{N-\alpha}(t+Ni+\alpha)}^2 
 \\
\label{eq:energypattern_lag_4} &=  \frac{2}{L-1} (1-\lambda)^{Ng} \av{E_{{}_{\lambda,\rho}}(J,\sigma)}   - \lambda     \sum_{i=0}^{g-1} (1-\lambda)^{gN-1-Ni}  \rho^{2Ni}\\ 
\label{eq:energypattern_lag_5} &= -\lambda \left( \frac{ (1-\lambda)^{N (g +1)} \rho^{2N}}{1- (1-\lambda)^{N} \rho^{2N}} + \frac{(1-\lambda)^{N (g+1) -1} - (1-\lambda)^{N-1} \rho^{2 N g}}{ (1-\lambda)^{N} - \rho^{2 N}}\right).
\end{align}
Here, we used the expression of the network's matrix $J$ in eq.~\ref{eq.Jn0_SI}  to  arrive at eq.~\ref{eq:energypattern_lag_2}, and then followed the procedure introduced in eq.~\ref{eq:J-with_patterns_second order} to arrived at the double-summation in eq.~\ref{eq:energypattern_lag_3}. We then used the equation for pattern overlap $\braket{\sigma^\alpha(t_1)|\sigma^\nu(t_2)} =  \delta_{\alpha,\nu} \rho^{|t_2-t_1|}$ to reduce the sum in eq.~\ref{eq:energypattern_lag_4} and  arrived at the result in eq.~\ref{eq:energypattern_lag_5} by evaluating the geometric sum and substituting the empirical average for the energy $\av{E_{{}_{\lambda,\rho}}(J,\sigma)}$ from eq.~\ref{eq:energypattern_final}.

We probe this lagged memory by looking at the performance $\Q$ for patterns that are correctly associated with  their memory attractors (i.e., those with  $\braket{\sigma_\text{att}|\sigma} >0.8$). As shown in Fig.~\ref{fig:Simresults_overleps_high}, for a broad parameter regime, the mean performance for these correctly associated patterns agrees well with the theoretical expectation  $\Q_\lag=  \rho^{g_\lag N}$, which is lower than the naive expectation $\Q_0$.

\newpage{}
\section{Structure of the energy landscape for working memory}
\renewcommand{\thesubsection}{C\arabic{subsection}}

\subsection{Formation of mountain passes in the energy landscape of memory for evolving patterns} 
As shown in Fig.~\ref{fig:Fig1}, large learning rates in networks with memory for evolving patterns result in the emergence of  narrow connecting paths between the  minima of the energy landscape. We refer to these narrow connecting paths  as  \textit{mountain passes}.  In pattern retrieval,  the Monte Carlo search can  drive a pattern out of one energy minimum  into another  minimum and potentially lead to pattern misclassification.

We use two features of the energy landscape to probe the emergence  of the mountain passes.
 
First, we show that if a pattern is misclassified, it has fallen into a memory attractor associated with another pattern class and not a spuriously made energy minima. To do so, we compare the overlap of the attractor with the original pattern $|\braket{\sigma^\alpha_{\text{att}}|\sigma^\alpha}|$ (i.e., the reconstruction performance of the patterns) with the maximal overlap of the attractor with all other patterns $\max_{\nu \neq \alpha}|\braket{\sigma^\alpha_{\text{att}}|\sigma^\nu}|$. Indeed, as shown in  Fig.~\ref{fig:Dist_dist_paths}A for evolving patterns, the  memory attractors associated with $99.4 \%$ of the originally stored patterns have either a large overlap with the correct pattern or with one of the other previously presented pattern. 71.3\% of  the patterns are correctly classified (stable patterns in sector {\bf I} in Fig.~\ref{fig:Dist_dist_paths}A), whereas 28.1\% of them are associated with a secondary energy minima after equilibration (unstable patterns in sector {\bf II} in Fig.~\ref{fig:Dist_dist_paths}A). A very small fraction of patterns ($<1\%$) fall into local minima given by the linear combinations of the presented patterns (sector {\bf IV} in Fig.~\ref{fig:Dist_dist_paths}A). These minima are well-known in the classical Hopfield model~\cite{Amit:1985kd,Fontanar:1990pf}. Moreover, we see that equilibration of a random pattern (i.e., a pattern orthogonal to all the presented classes) in the energy landscape leads to memory attractors for one of the originally presented pattern classes. The majority of these random patterns lie in sector  {\bf II} of Fig.~\ref{fig:Dist_dist_paths}A), i.e., they have a small overlap with   the network since they are orthogonal to the originally presented pattern classes, and they fall into one of the existing memory attractors after equilibration. 

Second, we characterize the possible paths for a pattern to move from one valley to another during equilibration, using Monte Carlo algorithm with the Metropolis acceptance probability,
\EQ
\rho(\sigma \to \sigma') = \text{min}\left(1, e^{-\beta( E(J,\sigma')-E(J,\sigma)) }\right)\EE

We estimate the number of beneficial spin-flips (i.e., open paths) that decrease the energy of a pattern at the start of equilibration (Fig.~\ref{fig:Dist_dist_paths}B). The average  number of open paths is smaller  for stable  patterns compared to the unstable patterns that are be miscalssified during retrieval (Fig.~\ref{fig:Dist_dist_paths}B). However, the distributions for the number of open paths largely overlap for stable and unstable patterns.  Therefore, the local energy landscape of stable and unstable patterns are quite  similar and it is difficult to discriminate between them solely based on the local gradients in the landscape. Fig.~\ref{fig:Paths_partisipation}A  shows that the average number of beneficial spin-flips grows with the mutation rate of the patterns but this number is comparable for stable and unstable patterns. Moreover, the unstable stored patterns (blue) have far fewer open paths available to them during equilibration compared to random patterns (red) that are presented to the network for the first time (Figs.~\ref{fig:Dist_dist_paths}B~\&~\ref{fig:Paths_partisipation}A).  Notably,  on average half of the spin-flips reduce the energy of for random patterns, irrespective of the mutation rate. This indicates that even though previously presented pattern classes are statistically distinct from random patterns,  they can still become  unstable, especially in networks which are presented with evolving patterns.

It should be noted that the evolution of the patterns only indirectly contribute to the misclassification of memory, as it necessitates a larger learning rate for the networks to optimally operate, which in turn results in the emergence of mountain passes. To clearly demonstrate this effect, Figs.~\ref{fig:Dist_dist_paths}C,D, and~\ref{fig:Paths_partisipation}D shows the  misclassification behavior for a network trained to store memory for static pattern, while using a larger learning rate that is optimized for  evolving patterns. Indeed, we see that pattern misclassification in this case is consistent with the existence of mountain passes in the network's energy  landscape.

\subsection{Spectral decomposition of the energy landscape}
We use spectral decomposition of the energy landscape to characterize the relative positioning and the stability of patterns in the landscape. 
As shown in Figs.~\ref{fig:Dist_dist_paths},~\ref{fig:Paths_partisipation}, destabilization of patterns due to equilibration over mountain passes occurs in networks with high learning rates, even for static patterns. Therefore, we focus on how the learning rate impacts the spectral decomposition of the  energy landscape in networks presented with static patterns. This simplification will enable us to analytically probe the structure of the energy landscape, which we will compare with numerical results for evolving patterns.

We can represent the network $J$ (of size $L\times L$) that store a memory of $N$ static patterns with $N$ non-trivial eigenvectors $\ket{\Phi^i}$ with corresponding eigenvalues $\Gamma_i$, and $N-L$ degenerate eigenvectors, $\ket{\Psi^k}$ with corresponding trivial eigenvalues $\gamma_k =\gamma=-1$:
\begin{align}
\label{eq:J_through_Gamma_full}
J= \sum_{i=1}^N \Gamma_i \ket{\Phi^i}\bra{\Phi^i} + \sum_{k=1}^{L-N} \gamma_k \ket{\Psi^k}\bra{\Psi^k}.
\end{align} 

The non-trivial  eigenvectors span the space of the presented patterns, for which the recognition energy can be expressed by
\begin{align}
\label{eq:energies_sved_patterns}
E(J,\sigma^\alpha)= -\frac{1}{2}\sum_{i=1}^N \Gamma_i \braket{{\sigma}^\alpha| \Phi^i}\braket{\Phi^i| {\sigma}^\alpha}.
\end{align}

An arbitrary configuration $\chi$ in general can have components orthogonal to the $N$ eigenvectors $\ket{\Phi^i}$, as it points to a vertex of the hypercube, and should be expressed in terms of all the eigenvectors $\{\Phi^1,\dots,\Phi^N,\Psi^1,\dots,\Psi^{L-N}\}$:
\begin{align}
\label{eq:energies_sved_patterns_full}
E(J,\chi)= -\frac{1}{2}\left(\underbrace{\sum_{i=1}^N \Gamma_i \braket{ {\chi}| \Phi^i}\braket{\Phi^i| {\chi}}}_{\text{stored patterns}} + \underbrace{\sum_{k=1}^{L-N} \gamma\braket{ {\chi}| \Psi^k}\braket{\Psi^k| {\chi}}}_{\text{trivial space}}\right).
\end{align}
 
Any spin-flip in a pattern (e.g., during equilibration) can be understood as a rotation in the eigenspace of the network (eq.~\ref{eq:energies_sved_patterns_full}). As a first step in characterizing these rotations we remind ourselves of the identity
\begin{align}
\ket{ {\chi}} = \sum_{i=1}^N \braket{\Phi^i| {\chi}} \ket{\Phi^i} + \sum_{k=1}^{L-N} \braket{\Psi^k| {\chi}} \ket{\Psi^k},
\end{align}
with the normalization condition
\begin{align}
\label{eq:Normed_overlaps}
\sum_{i=1}^N \left(\braket{\Phi^i| {\chi}}\right)^2 + \sum_{k=1}^{L-N} \left(\braket{\Psi^k| {\chi}}\right)^2 =1.
\end{align}
 In addition, since the diagonal elements of the network are set to $J_{ii}=0$ (eq.~\ref{eq:learningrule_SI}), the eigenvalues should sum to zero, or alternatively,
 \EQ 
 \sum_{i=1}^N \Gamma_i  =  - \sum_{k=1}^{L-N}  \gamma_k=  L - N.
 \EE

To asses the stability of a pattern ${\sigma}^\nu$,  we  compare its recognition  energy $E(J,{\sigma}^\nu)$ with the energy of the  rotated pattern after a spin-flip $E(J,\tilde{\sigma}^\nu)$. To do so, we first consider a simple scenario, where we assume that the pattern $\sigma^\nu$ has a large overlap with one {dominant} non-trivial  eigenvector $\Phi^A$ (i.e., $\braket{\sigma^\nu|\Phi^A}^2=m^2 \approx 1$). The other components of the pattern can be expressed in terms of  the remaining $N-1$ non-trivial eigenvectors with mean squared overlap $\frac{1-m^2}{N-1}$. The expansion of the recognition energy for the presented pattern is restricted to the $N$ non-trivial directions (eq.~\ref{eq:energies_sved_patterns_full}), resulting in
\begin{align}
\label{eq:enrgy_p1}
E(J,{\sigma}^\nu)=-\frac{1}{2}\left( m^2 \Gamma_A + \sum_{i\neq A} \frac{1-m^2}{N-1} \Gamma_i \right)= -\frac{1}{2}\left(  m^2 \Gamma_A  +(1-m^2) \tilde{\Gamma} \right),
\end{align}
where $\tilde{\Gamma} = \frac{1}{N-1}\sum_{i\neq A} \Gamma_i = \frac{N \bar{\Gamma} - \Gamma_A}{N-1}$ is the mean eigenvalue for the non-dominant directions. 

A spin-flip ($\ket{\sigma^\nu} \rightarrow \ket{\tilde{\sigma}^\nu}$ ) can rotate the pattern out of the dominant direction $\Phi^A$ and reduce the squared overlap by $\epsilon^2$. The rotated pattern $\ket{\tilde{\sigma}^\nu}$ in general lies in the $L$-dimensional space and is not restricted to the $N$-dimensional (non-trivial) subspace. We first take a  {\em mean-field} approach in describing the rotation of the pattern after a spin-flip. Because of the normalization condition (eq.~\ref{eq:Normed_overlaps}), the loss in the overlap with the dominant direction should result in {\em an average} increase in the overlap with  the other $L-1$  eigenvectors  by  $\frac{\epsilon^2}{L-1}$. The energy of the rotated pattern after a spin-flip $E(J,\tilde{\sigma}^\nu)$ can be expressed in terms of all the $L$ eigenvectors (eq.~\ref{eq:energies_sved_patterns_full}), 
\begin{align}
\nonumber E(J,\tilde{\sigma}^\nu) &= -\frac{1}{2}\left[ (m^2 - \epsilon^2) \Gamma_A + \sum_{i\neq A} \left( \frac{1-m^2}{N-1} + \frac{\epsilon^2}{L-1} \right) \Gamma_i  + \sum_{k} \frac{\epsilon^2}{L-1} \gamma_k \right] \\
 &= E(J,{\sigma}^\nu) +\frac{\epsilon^2}{2}\left[  \Gamma_A -\frac{1}{L-1} \left(\sum_{i\neq A} \Gamma_i  +\sum_{k}\gamma_k\right)\right]\label{eq:enrgy_p1}.\\
&=E(J,{\sigma}^\nu) +\frac{\epsilon^2}{2} \Gamma_A\left(  1 +\frac{ 1}{L-1} \right)\label{eq:enrgy_p2}.
\end{align}
where in eq.~\ref{eq:enrgy_p2} we used the fact that the eigenvalues should sum up to zero. On average, a spin-flip {$\ket{{\sigma}^\nu} \rightarrow \ket{\tilde{\sigma}^\nu}$} increases the recognition energy by $ E(J,\tilde{\sigma}^\nu)- E(J,{\sigma}^\nu)= \frac{\epsilon^2}{2} \Gamma_A\left[ 1 +\mathcal{O}\left(L^{-1}\right)\right]$.  This is consistent with the results shown in Figs.~\ref{fig:Dist_dist_paths}B,D and Figs.~\ref{fig:Paths_partisipation}A,D, which indicate that  the majority of the spin-flips keep a pattern in the original energy minimum and only  a few of the spin-flips drive a pattern out of the original attractor. 

In the analysis above, we assumed that the reduction in overlap with the dominant eigenvector $\epsilon^2$ is absorbed equally by all the other eigenvectors (i.e., the mean-field approach). 
In this case, the change in energy is equally distributed across the positive and the negative eigenvalues ($\Gamma$'s and $\gamma$'s in~eq.~\ref{eq:enrgy_p1}), resulting in an overall increase in the energy due to the reduced overlap with the dominant direction $\ket{\Phi^A}$. The destabilizing spin-flips are associated with  atypical changes that rotate a pattern onto  a secondary non-trivial direction $\ket{\Phi^B}$ (with positive eigenvalue $\Gamma_B$), as a result of which the total energy could be reduced. To better characterize the conditions under which patterns become unstable, we will introduce a perturbation to the mean-field approach used in eq.~\ref{eq:enrgy_p2}. We will assume that a spin-flip results in a rotation with a dominant component along a secondary non-trivial direction $\ket{\Phi^B}$. Specifically,  we will assume the reduced overlap $\epsilon^2$ between the original pattern $\ket{{\sigma}^\nu}$ and the dominant direction $\ket{\Phi^A}$ is distributed in an imbalanced fashion between the other eigenvectors, with a fraction  $p$  projected onto a new (non-trivial) direction  $\ket{\Phi^B}$,  while all the other $L-2$ directions span the remaining $(1-p)\epsilon^2$. In this case, the energy of the rotated pattern is given by
\begin{align}
\label{eq:enrgy_pp1}
\nonumber E(J,\tilde{\sigma}^\nu) &= -\frac{1}{2}\left[ (m^2 - \epsilon^2) \Gamma_A + \left( \frac{1-m^2}{N-1} + p\epsilon^2 \right) \Gamma_B +  \sum_{i\neq A,B} \left( \frac{1-m^2}{N-1} + \frac{(1-p)\epsilon^2}{L-2} \right) \Gamma_i  + \sum_{k} \frac{(1-p)\epsilon^2}{L-2} \gamma_k \right] \\
&=E(J,{\sigma}^\nu) +\frac{\epsilon^2 }{2}\left[\Gamma_A  -  p\, \Gamma_B   +\mathcal{O}\left(L^{-1}\right) \right].
\end{align}
Therefore, a spin-flip is beneficial if $ \Gamma_A  <  p\, \Gamma_B$.  To further concretize this condition, we will estimate the typical loss $\epsilon^2$ and gain $p\epsilon^2$ in the squared overlap between the pattern and its dominating directions due to rotation by a single spin-flip. 
 
Let us consider a rotation $\ket{\sigma^\nu} \to \ket{\tilde \sigma^\nu}$ by a flip in the $n^{th}$ spin of the original pattern $\ket{\sigma^\nu}$. This spin flip reduces the original overlap of the pattern $m=\braket{\sigma^\nu|\Phi^A}$ with the dominant direction $\ket{\Phi^A}$ by  the amount $\frac{2}{\sqrt{L}} \Phi_n^A$, where   $\Phi_n^A$ is the $n^{th}$ entry of the eigenvector $ \ket{\Phi^A}$. Since the original overlap is large (i.e., $m\simeq 1$), all entries of the dominant eigenvector are approximately $ \Phi_i^A \simeq {1}/{\sqrt L}, \, \forall i$, resulting in a reduced overlap of the rotated pattern $\braket{\sigma^\nu|\Phi^A} = m-\frac{2}{L}$. Therefore, the loss in the  squared overlap $\epsilon^2$ by a spin flip is given by
\begin{align}
\label{eq:epsilon_squared}
\epsilon^2= \braket{{\sigma}^\nu|\Phi^j}^2-\braket{\tilde{\sigma}^\nu|\Phi^j}^2 = m^2 - \left(m^2 - 4 \frac{m}{L} + 4 \frac{1}{L^2}\right) =  4 \frac{m}{L} +\mathcal{O}(\frac{1}{L^2}).
\end{align} 

Similarly, we  can derive the gain in the  squared overlap $p\epsilon^2$ between the pattern $\ket{\sigma^\nu}$  and the new  dominant direction $\ket{\Phi^B}$ after a spin-flip. Except for the direction $\ket{\Phi^A}$, the expected squared overlap between the original pattern (prior to the spin flip) and any of the non-trivial eigenvectors including $\ket{\Phi^B}$ is $ \braket{{\sigma}^\nu|\Phi^B}^2=\frac{1-m^2}{N-1}$. The  flip in the $n$-th spin of the original pattern increases the overlap of the rotated pattern with the new dominant direction $\ket{\Phi^B}$ by $2\frac{\Phi^B_n}{\sqrt{L}}$, where $\Phi^B_n$ should be of the order of $\sqrt{1/L}$. Therefore, the largest gain in overlap due to a spin-flip is given by
\begin{equation}
\begin{split}
p \epsilon^2= \braket{\tilde{\sigma}^\nu|\Phi^B}^2 -\braket{{\sigma}^\nu|\Phi^B}^2 &\simeq\left(\frac{1-m^2}{N-1} + 4 \sqrt{\frac{1-m^2}{N-1}} \frac{\Phi^B_n}{\sqrt{L}} + 4 \frac{(\Phi^B_n)^2}{L} \right) -\frac{1-m^2}{N-1} \\
\label{eq:epsilon_squared_p}
&=\sqrt{\frac{1-m^2}{N-1}} \frac{\Phi^B_n}{\sqrt{L}} +\mathcal{O}(\frac{1}{L^2}).
\end{split}
\end{equation} 
By using the results from eqs.~\ref{eq:epsilon_squared} and \ref{eq:epsilon_squared_p}, we can express the condition for beneficial spin-flips to drive a pattern over the carved mountain passes during equilibration (eq.~\ref{eq:enrgy_pp1}), 
\begin{align}
\epsilon^2 \Gamma_A  <  \epsilon^2 p \Gamma_B,\qquad \longrightarrow\qquad
\frac{\Gamma_A}{\Gamma_B}<\sqrt{\frac{1-m^2}{m^2}} \frac{1}{\sqrt{\alpha}} \Phi^B_n,
\label{eq:transition_condition_final}
\end{align}
where $\alpha={N}/{L}$. This result suggests that the  stability of a pattern depends on  how the ratio of the eigenvalues associated with the dominant projections of the pattern before and after the spin-flip $\Gamma_A/\Gamma_B$ compare to the  overlap $m$ of the original pattern with the dominant eigenvector $\Phi^A$ and the change due to  the spin-flip $\Phi_n^B$.

So far, we have constrained our analysis to patterns that have a dominant contribution to only one eigenvector $\Phi^A$. To extend our analysis to patterns which are instead constrained to a small sub-space $\mathcal{A}$ spanned by non-trivial eigenvalues, we define the squared pattern overlap with the subspace  $m_\mathcal{A}^2 = \sum_{a \in \mathcal{A}}\braket{\sigma^\nu|\Phi^a}^2$ and a weighted averaged eigenvalue in the subspace $\Gamma_\mathcal{A} =  \sum_{a \in \mathcal{A}}\braket{\sigma^\nu|\Phi^a}^2 \Gamma_a$. 
As a result, the  difference in the energy of a pattern before and after a spin-flip (eq.~\ref{eq:enrgy_pp1}) can be extended to $E(J,{\sigma}^\nu) -E(J,\tilde{\sigma}^\nu) =\frac{\epsilon^2 }{2}\left[\Gamma_\mathcal{A}  -  p\, \Gamma_B   +\mathcal{O}\left(L^{-1}\right) \right]$. Similarly, the stability condition   in eq.~\ref{eq:transition_condition_final} can be extended to $\frac{\Gamma_\A}{\Gamma_B}<\sqrt{\frac{1-m_\A^2}{m_\A^2}} \frac{1}{\sqrt{\alpha}} \Phi^B_n$.
Patterns that are constrained to larger subspaces are more stable, since the weighted averaged eigenvalue for their containing subspace  $\Gamma_\A$ is closer to the mean of all eigenvalues $\bar{\Gamma} = 1-N/L$ (law of large numbers). Therefore, in these cases a much larger eigenvalue gap (or a broader eigenvalue spectrum) is necessary to satisfy the condition for pattern instability.

Fig.~\ref{fig:Scatter_current_next} compares the loss  in energy with the original dominant direction $\epsilon^2 \Gamma_A$ to the maximal gain in any of the other directions  $\epsilon^2 p \Gamma_B $ to test the pattern stability criteria presented in eq.~\ref{eq:transition_condition_final}. To do so, we identify a spin flip  $n$ in a secondary  direction $B$ that confers the maximal energy gain: $\epsilon^2 p \Gamma_B  \approx \max_{n,B} \sqrt{\frac{1-m^2}{N-1}} \frac{\Phi^B_n}{\sqrt{L}}\Gamma_B$. In Fig.~\ref{fig:Scatter_current_next}A,C we specifically focus on the subset of patterns that show a large (squared) overlap with the one dominant direction (i.e.,  $m > 0.85$). Given that evolving patterns are not constraint to the $\{\Phi \}$ (non-trivial) sub-space, we find a smaller fraction of these  patterns to fulfill the condition for such large overlap $m$ (Fig.~\ref{fig:Scatter_current_next}A), compared to the static patterns (Fig.~\ref{fig:Scatter_current_next}C). Nonetheless, we see that the criteria in eq.~\ref{eq:transition_condition_final} can be used to predict the stability of patterns in a network for both static and evolving patterns; note that here we use the same learning rate for both the static and evolving patterns.

We then relax the overlap condition by  including all patterns that have a large overlap with a subspace $\A$, spanned by up to 10 eigenvectors (i.e., $m_\A^2 = \sum_{\alpha\in\A} \braket{\sigma|\Phi^\alpha}^2>0.85$). For these larger subspaces the transition between stable and unstable patterns is no longer exactly given by eq.~\ref{eq:transition_condition_final}. However, the two contributions $\epsilon^2 \Gamma_\A$ and $\epsilon^2 p \Gamma_B$ still clearly separate the patterns into stable and unstable classes for both static and evolving patterns (Figs.~\ref{fig:Scatter_current_next}B,D). The softening of this condition is expected as in this regime we can no longer assume that a single spin-flip can reduce the overlap with all the eigenvectors in the original subspace. As a result, the effective loss in overlap become  smaller than $\epsilon^2$ and patterns become unstable below the dotted line in Fig.~\ref{fig:Scatter_current_next}B,D.

As the learning rate increases, the gap between the eigenvalues $\Gamma_B/\Gamma_A$ (Fig.~\ref{fig:eigenvalues}) become larger. At the same time,   patterns become more constrained to smaller subspaces (Fig.~\ref{fig:Paths_partisipation}C,D). As a result of these two effects, more patterns satisfy the instability criteria in eq.~\ref{eq:transition_condition_final}. These patterns  are misclassified as they fall into a wrong energy minimum by equilibrating through the mountain passes carved in  the energy landscape  of networks with large learning rates.

%%------------------------

%\bibliography{armita}
%\bibliography{workingMem,armita}

\begin{thebibliography}{10}
\providecommand{\url}[1]{\texttt{#1}}
\providecommand{\urlprefix}{URL }
\expandafter\ifx\csname urlstyle\endcsname\relax
  \providecommand{\doi}[1]{doi:\discretionary{}{}{}#1}\else
  \providecommand{\doi}{doi:\discretionary{}{}{}\begingroup
  \urlstyle{rm}\Url}\fi
\providecommand{\bibAnnoteFile}[1]{%
  \IfFileExists{#1}{\begin{quotation}\noindent\textsc{Key:} #1\\
  \textsc{Annotation:}\ \input{#1}\end{quotation}}{}}
\providecommand{\bibAnnote}[2]{%
  \begin{quotation}\noindent\textsc{Key:} #1\\
  \textsc{Annotation:}\ #2\end{quotation}}
\providecommand{\eprint}[2][]{\url{#2}}

\bibitem{Labrie:2010bc}
Labrie SJ, Samson JE, Moineau S (2010) {Bacteriophage resistance mechanisms}.
\newblock Nature Rev Microbiol 8: 317--327.
\input{Labrie:2010bc}

\bibitem{Barrangou:2014ht}
Barrangou R, Marraffini LA (2014) {CRISPR-Cas systems: Prokaryotes upgrade to
  adaptive immunity}.
\newblock Molecular cell 54: 234--244.
\input{Barrangou:2014ht}

\bibitem{Bradde:2020kb}
Bradde S, Nourmohammad A, Goyal S, Balasubramanian V (2020) {The size of the
  immune repertoire of bacteria}.
\newblock Proc Natl Acad Sci USA 117: 5144--5151.
\input{Bradde:2020kb}

\bibitem{Perelson:1997dla}
Perelson AS, Weisbuch G (1997) {Immunology for physicists}.
\newblock Rev Mod Phys 69: 1219--1267.
\input{Perelson:1997dla}

\bibitem{Janeway:2001te}
Janeway C, Travers P, Walport M, Schlomchik M (2001) {Immunobiology}.
\newblock The Immune System in Health and Disease. New York: Garland Science, 5
  edition.
\input{Janeway:2001te}

\bibitem{AltanBonnet:2020hk}
Altan-Bonnet G, Mora T, Walczak AM (2020) {Quantitative immunology for
  physicists}.
\newblock Physics Reports 849: 1--83.
\input{AltanBonnet:2020hk}

\bibitem{Haberly:1989jo}
Haberly LB, Bower JM (1989) {Olfactory cortex: Model circuit for study of
  associative memory?}
\newblock Trends Neurosci 12: 258--264.
\input{Haberly:1989jo}

\bibitem{Brennan:1990bp}
Brennan P, Kaba H, Keverne EB (1990) {Olfactory recognition: A simple memory
  system}.
\newblock Science 250: 1223--1226.
\input{Brennan:1990bp}

\bibitem{Granger:1991hu}
Granger R, Lynch G (1991) {Higher olfactory processes: Perceptual learning and
  memory}.
\newblock Curr Opin Neurobiol 1: 209--214.
\input{Granger:1991hu}

\bibitem{Haberly:2001gx}
Haberly LB (2001) Parallel-distributed processing in olfactory cortex: {N}ew
  insights from morphological and physiological analysis of neuronal circuitry.
\newblock Chemical senses 26: 551--576.
\input{Haberly:2001gx}

\bibitem{Wilson:2004do}
Wilson DA, Best AR, Sullivan RM (2004) {Plasticity in the olfactory system:
  {L}essons for the neurobiology of memory}.
\newblock Neuroscientist 10: 513--524.
\input{Wilson:2004do}

\bibitem{Raguso:2008re}
Raguso RA (2008) Wake {Up} and {Smell} the {Roses}: {The} {Ecology} and
  {Evolution} of {Floral} {Scent}.
\newblock Annu Rev Ecol Evol Syst 39: 549--569.
\input{Raguso:2008re}

\bibitem{Dunkel:2014na}
Dunkel A, et~al. (2014) Nature's chemical signatures in human olfaction: A
  foodborne perspective for future biotechnology.
\newblock Angew Chem Int Ed 53: 7124--7143.
\input{Dunkel:2014na}

\bibitem{Beyaert:14br}
Beyaert I, Hilker M (2014) Plant odour plumes as mediators of plant-insect
  interactions: {Plant} odour plumes.
\newblock Biol Rev 89: 68--81.
\input{Beyaert:14br}

\bibitem{glusman:2001gr}
Glusman G, Yanai I, Rubin I, Lancet D (2001) The {Complete} {Human} {Olfactory}
  {Subgenome}.
\newblock Genome Research 11: 685--702.
\input{glusman:2001gr}

\bibitem{Bargmann:2006na}
Bargmann CI (2006) Comparative chemosensation from receptors to ecology.
\newblock Nature 444: 295--301.
\input{Bargmann:2006na}

\bibitem{Touhara:2009AP}
Touhara K, Vosshall LB (2009) Sensing {Odorants} and {Pheromones} with
  {Chemosensory} {Receptors}.
\newblock Annu Rev Physiol 71: 307--332.
\input{Touhara:2009AP}

\bibitem{su:2009cl}
Su CY, Menuz K, Carlson JR (2009) Olfactory {Perception}: {Receptors}, {Cells},
  and {Circuits}.
\newblock Cell 139: 45--59.
\input{su:2009cl}

\bibitem{verbeurgt:2014po}
Verbeurgt C, et~al. (2014) Profiling of {Olfactory} {Receptor} {Gene}
  {Expression} in {Whole} {Human} {Olfactory} {Mucosa}.
\newblock PLoS ONE 9: e96333.
\input{verbeurgt:2014po}

\bibitem{Shepherd:1998ur}
Shepherd GM, Greer CA (1998) {Olfactory bulb}.
\newblock In: The Synaptic Organization of the Brain, 4th ed., New York, NY,
  US: Oxford University Press. pp. 159--203.
\input{Shepherd:1998ur}

\bibitem{Bushdid:2014sc}
Bushdid C, Magnasco MO, Vosshall LB, Keller A (2014) Humans {Can}
  {Discriminate} {More} than 1 {Trillion} {Olfactory} {Stimuli}.
\newblock Science 343: 1370--1372.
\input{Bushdid:2014sc}

\bibitem{Gerkin:2015el}
Gerkin RC, Castro JB (2015) The number of olfactory stimuli that humans can
  discriminate is still unknown.
\newblock eLife 4: e08127.
\input{Gerkin:2015el}

\bibitem{Mayhew:2020bx}
Mayhew EJ, et~al. (2020) Drawing the {Borders} of {Olfactory} {Space}.
\newblock preprint, Neuroscience.
\newblock \doi{10.1101/2020.12.04.412254}.
\newblock
  \urlprefix\url{http://biorxiv.org/lookup/doi/10.1101/2020.12.04.412254}.
\input{Mayhew:2020bx}

\bibitem{Lansner:2009hb}
Lansner A (2009) {Associative memory models: From the cell-assembly theory to
  biophysically detailed cortex simulations}.
\newblock Trends Neurosci 32: 178--186.
\input{Lansner:2009hb}

\bibitem{Hebb:1949vs}
Hebb DO (1949) {The Organization of Behavior: A Neuropsychological Theory}.
\newblock New York: Wiley.
\input{Hebb:1949vs}

\bibitem{Hopfield:1982fq}
Hopfield JJ (1982) {Neural networks and physical systems with emergent
  collective computational abilities}.
\newblock Proc Natl Acad Sci USA 79: 2554--2558.
\input{Hopfield:1982fq}

\bibitem{Mayer:2015ce}
Mayer A, Balasubramanian V, Mora T, Walczak AM (2015) {How a well-adapted
  immune system is organized.}
\newblock Proc Natl Acad Sci USA 112: 5950--5955.
\input{Mayer:2015ce}

\bibitem{Shinnakasu:2016ei}
Shinnakasu R, et~al. (2016) {Regulated selection of germinal-center cells into
  the memory B cell compartment}.
\newblock Nat Immunol 17: 861--869.
\input{Shinnakasu:2016ei}

\bibitem{Shinnakasu:2017ct}
Shinnakasu R, Kurosaki T (2017) {Regulation of memory B and plasma cell
  differentiation}.
\newblock Curr Opin Immunol 45: 126--131.
\input{Shinnakasu:2017ct}

\bibitem{Mayer:2019is}
Mayer A, Balasubramanian V, Walczak AM, Mora T (2019) {How a well-adapting
  immune system remembers}.
\newblock Proc Natl Acad Sci USA 116: 8815--8823.
\input{Mayer:2019is}

\bibitem{Schnaack:2020vb}
Schnaack OH, Nourmohammad A (2021) Optimal evolutionary decision-making to
  store immune memory.
\newblock eLife 10: e61346.
\input{Schnaack:2020vb}

\bibitem{Viant:2020jt}
Viant C, et~al. (2020) Antibody affinity shapes the choice between memory and
  germinal center {B} cell fates.
\newblock Cell 183: 1298--1311.e11.
\input{Viant:2020jt}

\bibitem{workingMem87}
Mezard M, Nadal JP, Toulouse G (1986) Solvable models of working memories.
\newblock J Physique 47: 1457-1462.
\input{workingMem87}

\bibitem{Amit:1985bo}
Amit DJ, Gutfreund H, Sompolinsky H (1985) Storing infinite numbers of patterns
  in a spin-glass model of neural networks.
\newblock Phys Rev Lett 55: 1530--1533.
\input{Amit:1985bo}

\bibitem{McEliece:1987ie}
McEliece R, Posner E, Rodemich E, Venkatesh S (1987) The capacity of the
  hopfield associative memory.
\newblock IEEE Transactions on Information Theory 33: 461-482.
\input{McEliece:1987ie}

\bibitem{Bouchaud:1997pr}
Bouchaud JP, Mezard M (1997) Universality classes for extreme-value statistics.
\newblock J Phys A Math Gen 30: 7997--8016.
\input{Bouchaud:1997pr}

\bibitem{Derrida:1997rd}
Derrida B (1997) From random walks to spin glasses.
\newblock Physica D: Nonlinear Phenomena 107: 186--198.
\input{Derrida:1997rd}

\bibitem{Goodfellow-et-al-2016}
Goodfellow I, Bengio Y, Courville A (2016) Deep Learning.
\newblock MIT Press.
\newblock \url{http://www.deeplearningbook.org}.
\input{Goodfellow-et-al-2016}

\bibitem{Mehta:2019pr}
Mehta P, et~al. (2019) A high-bias, low-variance introduction to {Machine}
  {Learning} for physicists.
\newblock Physics Reports 810: 1--124.
\input{Mehta:2019pr}

\bibitem{parga:1986jp}
Parga N, Virasoro M (1986) The ultrametric organization of memories in a neural
  network.
\newblock J Phys France 47: 1857--1864.
\input{parga:1986jp}

\bibitem{Virasoro:1986gk}
Virasoro MA (1986) {Ultrametricity, Hopfield Model and all that}.
\newblock In: Disordered Systems and Biological Organization, Springer, Berlin,
  Heidelberg. pp. 197--204.
\input{Virasoro:1986gk}

\bibitem{gutfreund:1988pr}
Gutfreund H (1988) Neural networks with hierarchically correlated patterns.
\newblock Phys Rev A 37: 570--577.
\input{gutfreund:1988pr}

\bibitem{tsodyks:1990mp}
Tsodyks MV (1990) Hierarchical associative memory in neural networks with low
  activity level.
\newblock Mod Phys Lett B 04: 259--265.
\input{tsodyks:1990mp}

\bibitem{Amit:1985kd}
Amit DJ, Gutfreund H, Sompolinsky H (1985) Spin-glass models of neural
  networks.
\newblock Physical Review A 32: 1007--1018.
\input{Amit:1985kd}

\bibitem{Fontanar:1990pf}
Fontanari JF (1990) Generalization in a hopfield network.
\newblock J Phys France 51: 2421--2430.
\input{Fontanar:1990pf}

\end{thebibliography}
%\bibliographystyle{plos2}

\newpage
\counterwithin{figure}{section}

  \setcounter{figure}{0}
 \renewcommand{\thefigure}{S\arabic{figure}}

\noindent{\Large \bf Supplementary Figures}\\\\  \\
  \begin{figure}[h!]
\centering
\includegraphics[width=\textwidth]{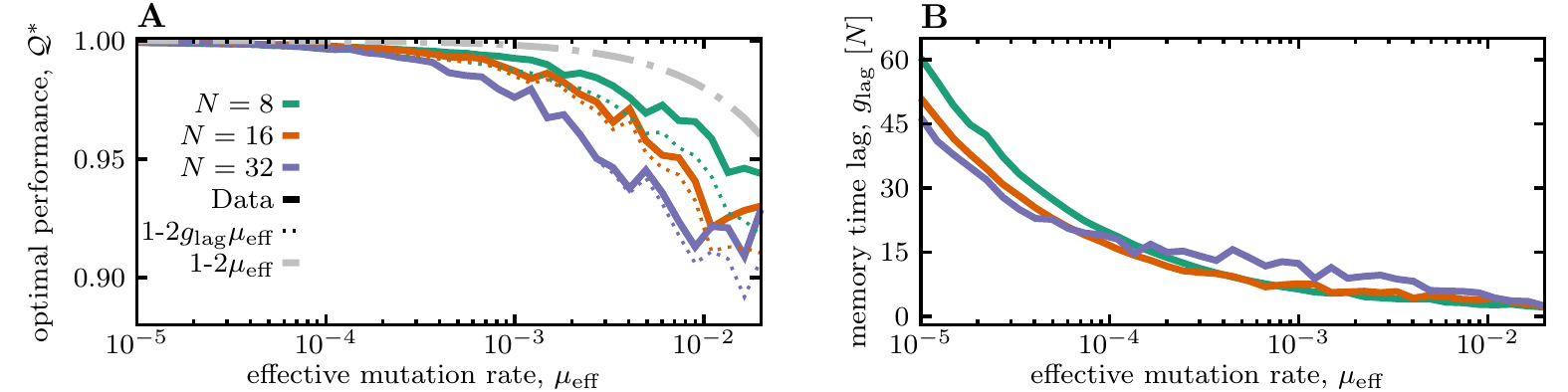}
\caption{\textbf{Reduced performance of Hopfield networks due to memory delay.} \textbf{(A)} The optimal performance  $\Q^*$ for patterns that are correctly associated with their memory attractors (i.e., they have an overlap $q(\sigma)=\braket{\sigma_\text{att}|\sigma} >0.8$) is shown as a function  of the effective mutation rate $\mu_\eff$. The solid lines show the simulation results for networks encountering a different number of patterns $N$ (colors). The gray dashed line shows the na\"ive expectation for the performance ($\Q_0= 1- 2\mu_\eff$), and the colored dashed lines show the expected performance after accounting for the memory lag $\Q_\lag = 1 -2 g_\lag \mu_\eff$. \textbf{(B)} The lag time $g_\lag$ for memory  is shown in units of generations [$N$] as a function of the effective mutation rate for networks encountering a different number of patterns (colors similar to (A)). The networks are trained with a learning rate $\lambda^*(\mu)$  optimized for the  mutation rate specified on the x-axis. Other simulation parameters: $L=800$.}
\label{fig:Simresults_overleps_high} 
\end{figure}

\vspace{3cm}

  \begin{figure}[h!]
\centering
\includegraphics[width=\textwidth]{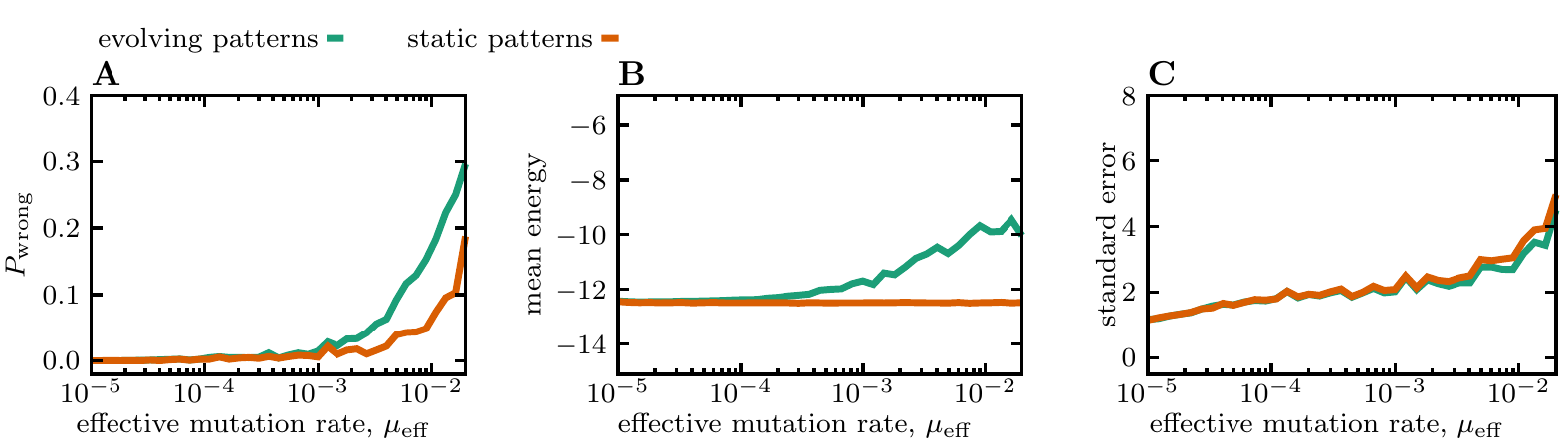}
 \caption{\textbf{Statistics of static and evolving patterns for networks with different learning rates.} 
  We compare the statistics of evolving (green) and static (orange) patterns in networks trained with a learning rate $\lambda^*(\mu)$  optimized for the  mutation rate specified on each panel's x-axis; see Fig.~\ref{Fig2}B for dependency of the optimal learning rate on mutation rate. The reported statistics are {\bf (A)} Fraction $P_{\text{wrong}}$ of misclassified  patterns  (i.e., patterns with a small overlap $q(\sigma) = \braket{\sigma_\text{att}|\sigma}<0.8$), {\bf (B)} the mean energy of the patterns, and \textbf{(C)} the standard error of the energy of the patterns in the network. Simulation parameters: $L=800$ and $N=32$. }
\label{fig:Performance_no_evo} 
\end{figure}

  \begin{figure}[h!]
\centering
\includegraphics[width=0.49\textwidth]{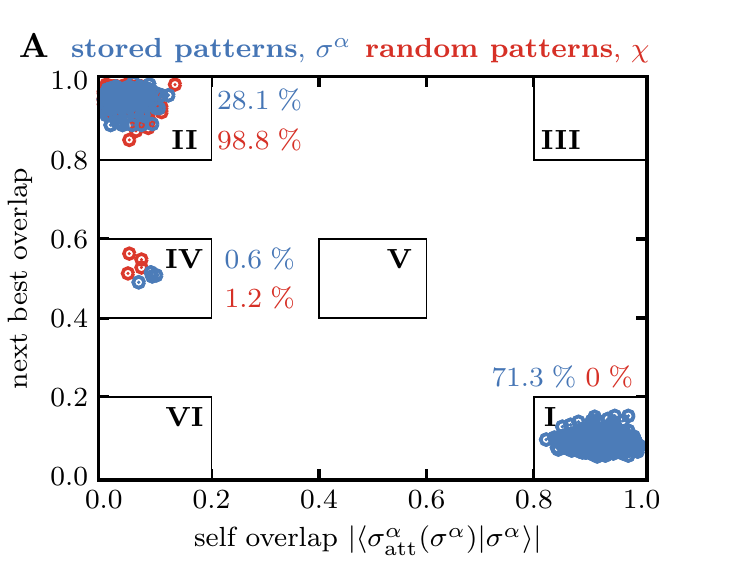}
\includegraphics[width=0.49\textwidth]{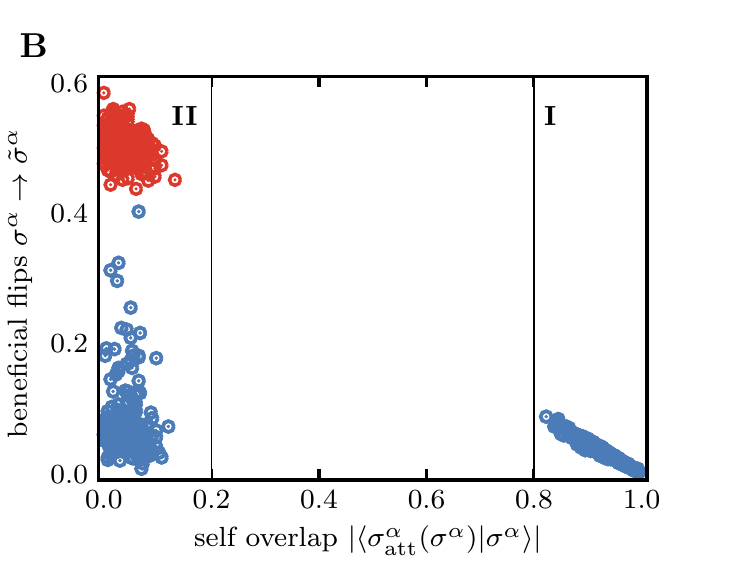}
\includegraphics[width=0.49\textwidth]{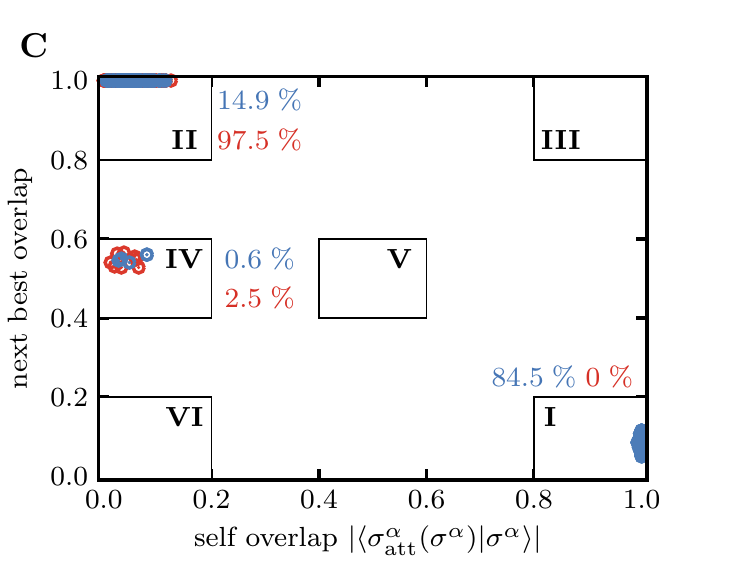}
\includegraphics[width=0.49\textwidth]{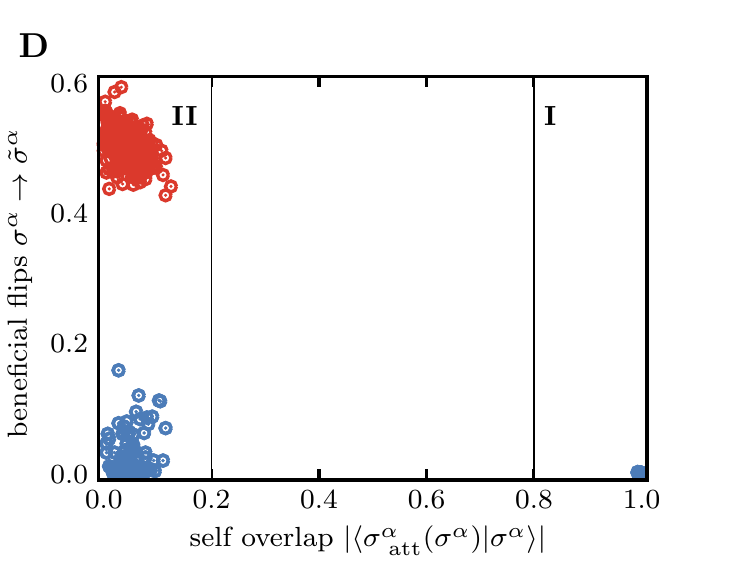}
 \caption{\textbf{Attractors and equilibration paths in networks.} Overlap of patterns with the networks' attractors are shown for  
 both the patterns  $\sigma^\alpha$ associated with one of the classes that were previously presented to the network during training (blue) and the random patterns $\chi$ that are on average orthogonal to the previously presented classes.
 {\bf(A)} The overlap between a presented pattern $\sigma^\alpha$ and the memory associated with the same pattern class $\sigma^\alpha_{\text{att}}(\sigma^{\alpha})$ is shown against the overlap of the pattern with the next best memory attractor associated with any of the other presented pattern classes $ \max_{\nu \neq\alpha} |\braket{\sigma^\alpha_\att|\sigma^\nu}|$. Fractions of the previously presented patterns and the random patterns that fall into different sectors of the plot are indicated in blue and red, respectively. 
Sector \textbf{I} corresponds to patterns that fall into the correct energy attractors (i.e., $\braket{\sigma^\alpha_{\text{att}}|\sigma^\alpha } \approx 1$). In the limit of large self-overlap, the maximal overlap to any other pattern family is close to zero, and thus, no patterns are found in sector \textbf{III}. Patters with a small self-overlap could fall into three different sectors: Sector \textbf{II} corresponds to misclassified patterns that fall into a valley associated with a different class ($\max_{\nu\neq \alpha}|\braket{\sigma^\alpha_{\text{att}} |\sigma^\nu }| \approx 1$). Patterns in sectors \textbf{IV} and \textbf{V} fall into local valleys between the minima of two pattern families. This   \textit{mixture states} are well known in the classical Hopfield model \cite{Amit:1985kd,Fontanar:1990pf}.  Sector \textbf{VI} indicates patterns that fall into an attractor in the network that does not correspond to any of the   previously presented classes. The fact that neither  the  previously presented patterns nor the random patterns fall into this sector suggests that the network indeed only stores memory of the presented patterns and is  not in the glassy regime. 
{\bf (B)} The number of beneficial spin-flips for presented pattern at the beginning of equilibration  (i.e., the number of open equilibration paths) is shown against pattens' self-overlap (x-axis in (A)). For  stable patterns (sector \textbf{I}) the number of open paths is anti-correlated with the overlap between the attractor and the presented pattern. For unstable patterns (sector \textbf{II}), the number of open paths is on average  larger that that of  the stable patterns. However, there are fewer paths available to the previously presented patterns compared to the random patterns.
In (A,B) patterns  evolve with rate $\mu_\eff = 0.01$ and the network's learning rate is optimized accordingly. 
The sharp transition between sector occupations indicates that our results are insensitive to the classification threshold for self-overlap (currently set to $q^\alpha>0.8$), i.e. any threshold value between sectors \textbf{I} and \textbf{II} would result in the same classification of patterns.  {\bf(C,D)} Similar to   (A,B) but for static patterns in a network with a similar learning rate to (A,B). 
Simulation parameters: $L=800$ and $N=32$.  
}
\label{fig:Dist_dist_paths} 
\end{figure}

  \begin{figure}[h!]
\centering
\includegraphics[width=\textwidth]{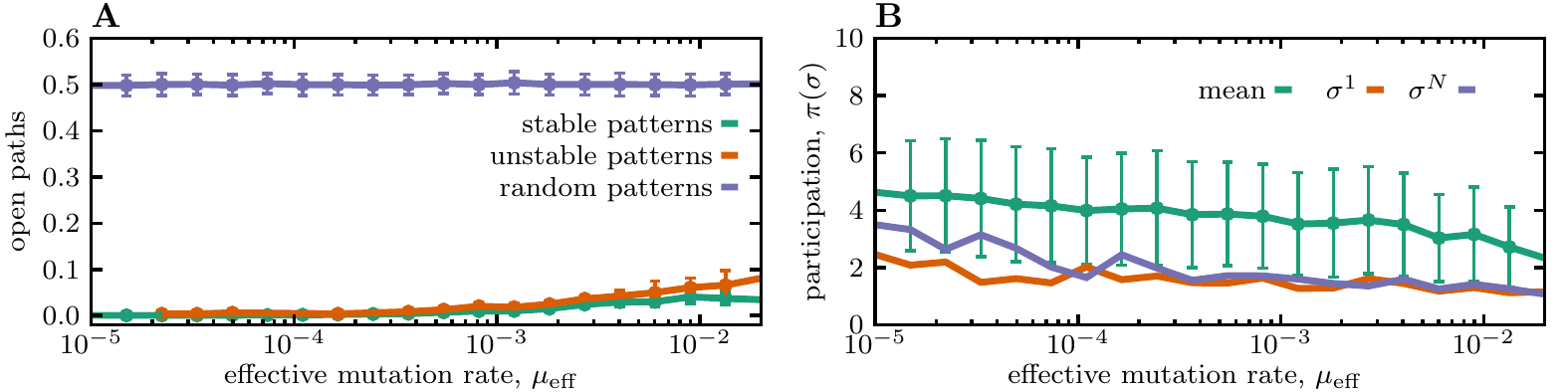}
\includegraphics[width=\textwidth]{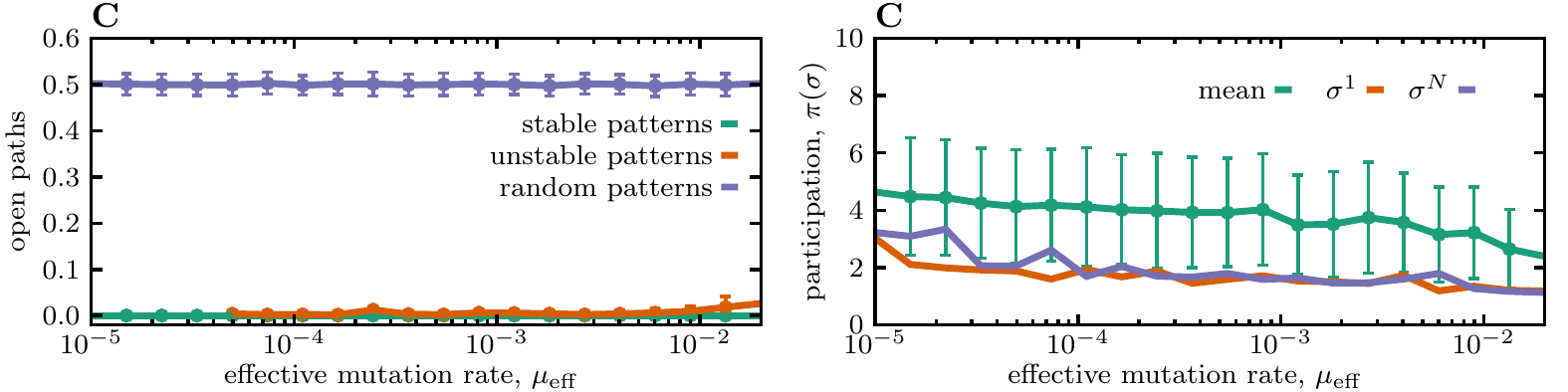}
 \caption{\textbf{Open equilibration paths and participation ratio.} (\textbf{A}) The mean number of open paths (i.e., the beneficial spin-flips at the beginning of equilibration)  is shown for stable, unstable, and random patterns (colors) as a function of the effective mutation rate $\mu_\eff$ in networks trained with the optimal learning rate  $\lambda^*(\mu)$. 
 (\textbf{B}) The participation ratio  $\pi(\sigma^j) = \frac{\left(\sum_i m_{i,j}^2\right)^2}{\sum_i m_{i,j}^4}$, with  $m_{i,j} = \braket{\Phi^i|\sigma^j}$ is shown for the pattern $\sigma^1$ with the lowest energy (orange), the  l pattern $\sigma^N$ with the highest energy (purple). The mean participation ratio averaged over all patterns is shown in green. 
 {\bf (C,D)} Similar to (A,B) but for static patterns ($\mu =0$). The learning rate of the network in this case is tuned to be optimal for the  mutation rate specified on the x-axis. Simulation parameters: $L=800$ and $N=32$. \\\\}
\label{fig:Paths_partisipation} 
\end{figure}

\vspace{3cm}

  \begin{figure}[h!]
\centering
\includegraphics[width=\textwidth]{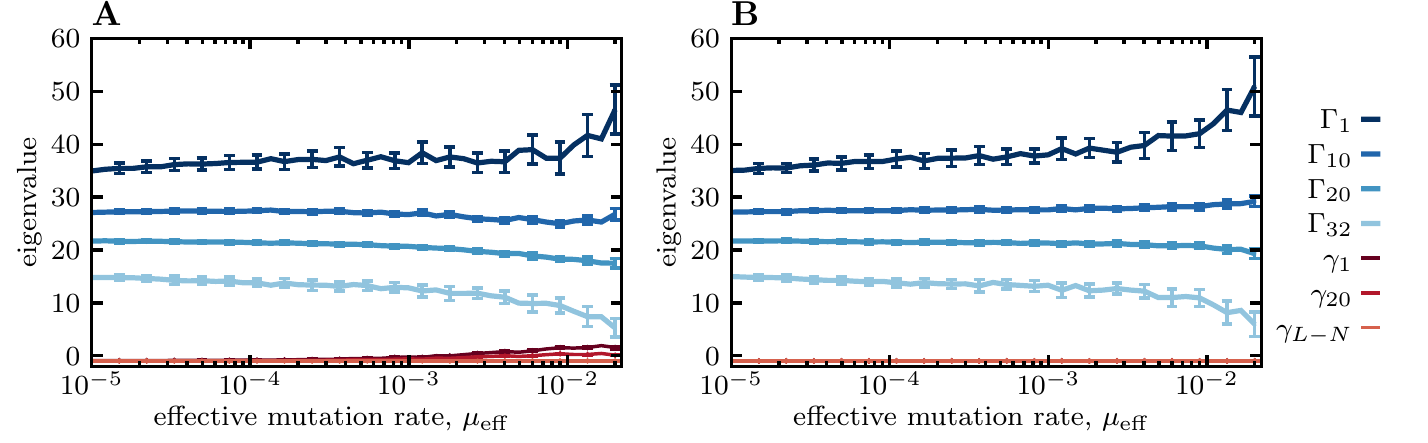}
 \caption{\textbf{Eigenvalues of networks with memory against dynamic and static patterns. }  (\textbf{A}) The  first $\Gamma_1$, the $10^{th}$   ($\Gamma_{10}$), the $20^{th}$ ($\Gamma_{20}$), and the last ($\Gamma_{N=32}$) non-trivial eigenvalues of a network of size $L=800$ presented with $N=32$ patterns is shown  as a function the patterns' effective mutation rate (different shades of blue). In each case, the network is trained with the optimal learning rate  $\lambda^*(\mu)$. 
 The trivial eigenvalues are shown in different shades of red, with their rank indicated in the legend. For small $\mu_\eff$ all trivial eigenvalues match the prediction  $\gamma_k=-1$, which implies that  the network updates fast enough to keep the patterns  within the $N$-dimensional sub-space. For larger mutation rates, some of the trivial eigenvalues  deviate from $-1$, indicating that evolving patterns start spanning in a larger sub-space.  Moreover, as the mutation rate (or learning rate) increases the gap between between the non-trivial eigenvalues broadens. {\bf (B)} Similar to (A) but for static patterns in networks trained with a learning rate $\lambda^*(\mu)$  optimized for the  mutation rate specified on the panel's x-axis.  In contrast to (A) all trivial eigenvalues remain equal to $-1$ independent of the learning rate, implying that the static patterns remained within the non-trivial $N$-dimensional sub-space.  Similar to (A) the gap between the nontrivial eigenvalues broadens with increasing learning rate.   }
\label{fig:eigenvalues} 
\end{figure}

  \begin{figure}[h!]
\centering
\includegraphics[width=0.49\textwidth]{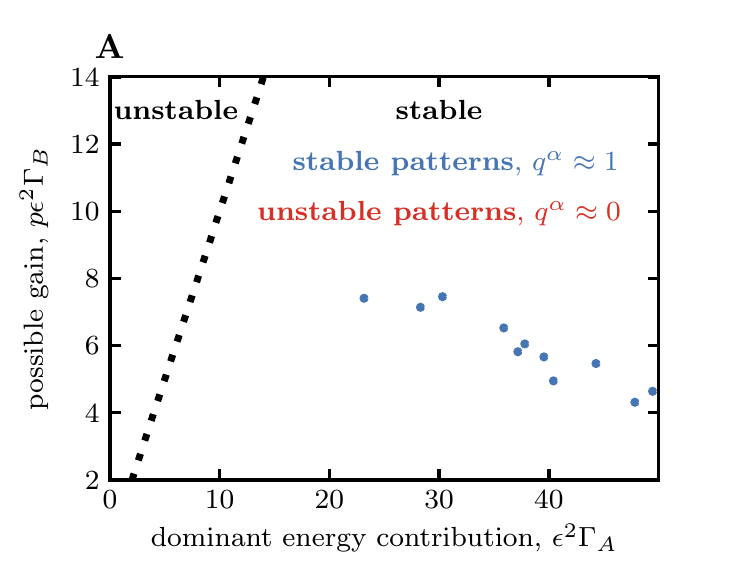}
\includegraphics[width=0.49\textwidth]{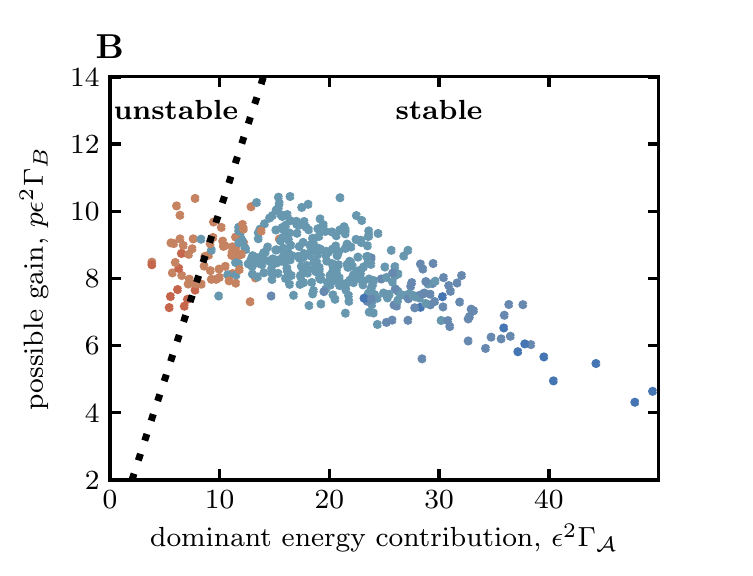}
\includegraphics[width=0.49\textwidth]{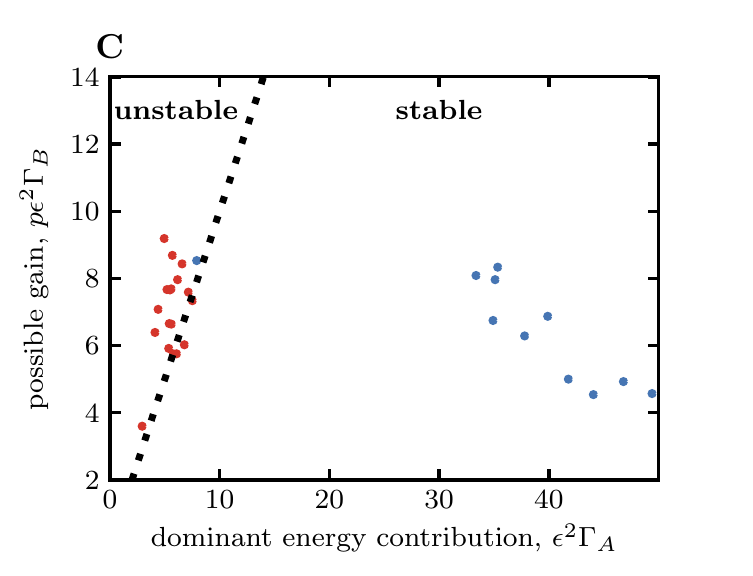}
\includegraphics[width=0.49\textwidth]{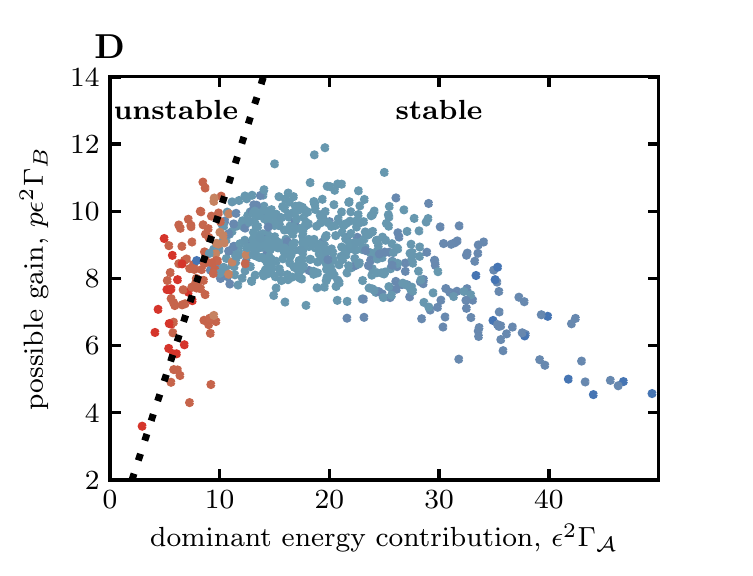}
 \caption{\textbf{Stability condition for patterns during equilibriation.} 
 The stability condition in eq.~\ref{eq:transition_condition_final} (dotted line) is used to classify stable (blue) and unstable (red) patterns for  \textbf{(A)} the patterns that have a  squared overlap with one dominant eigenvector $m^2=\braket{\Phi^A|\sigma^\nu}^2 >  0.85$, and {\bf(B)} the patterns that are constrained to a small sub-space $\mathcal{A}$ spanned by up to 10 nontrivial eigenvectors; in this case, $m_\mathcal{A}^2 =\sum_{a \in \mathcal{A} } \braket{\Phi^a|\sigma^\nu}^2 >0.85$.
The shading indicate the number of eigenvectors needed to represent a pattern from dark (one) to light (ten).  \textbf{(C, D)} Similar to (A, B)  but for static patterns in networks trained with the same learning rate as in  {(A, B)}. In general more static patterns reach the threshold of $m>0.85$ as these patterns remain constrained to the  N-dimensional subspace spanned by the non-trivial eigenvectors $\{\Phi_i\}$. Simulation parameters:  $N=32$,  $L=800$, $\mu_{\text{eff}} = 0.02$, and networks are trained with the optimal learning rate $\lambda^*(\mu)$.
}
\label{fig:Scatter_current_next} 
\end{figure}

\begin{figure}[h!]
\centering
\includegraphics[width=0.48\textwidth]{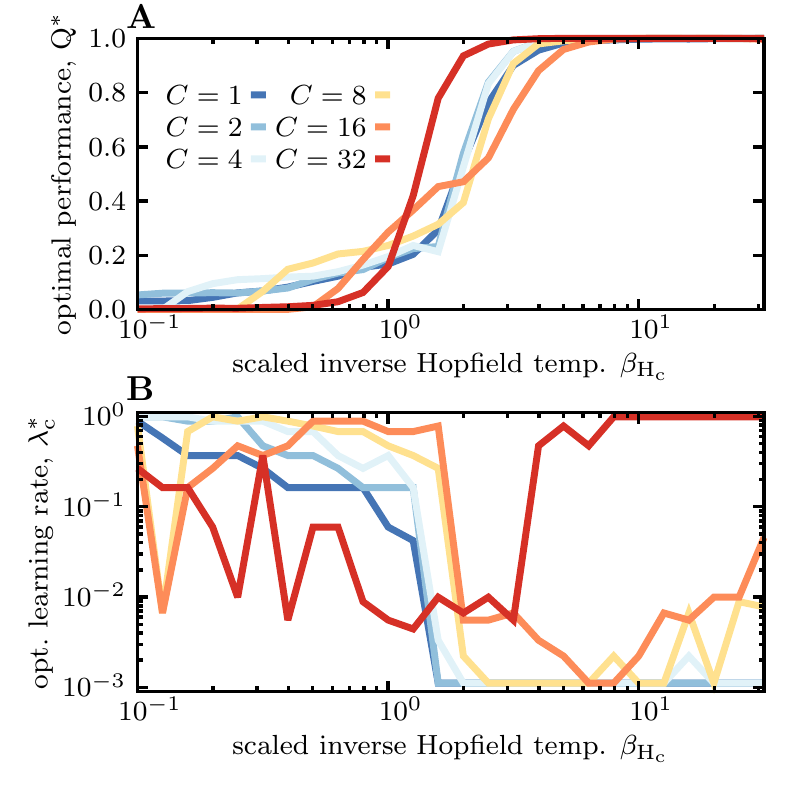}
\includegraphics[width=0.48\textwidth]{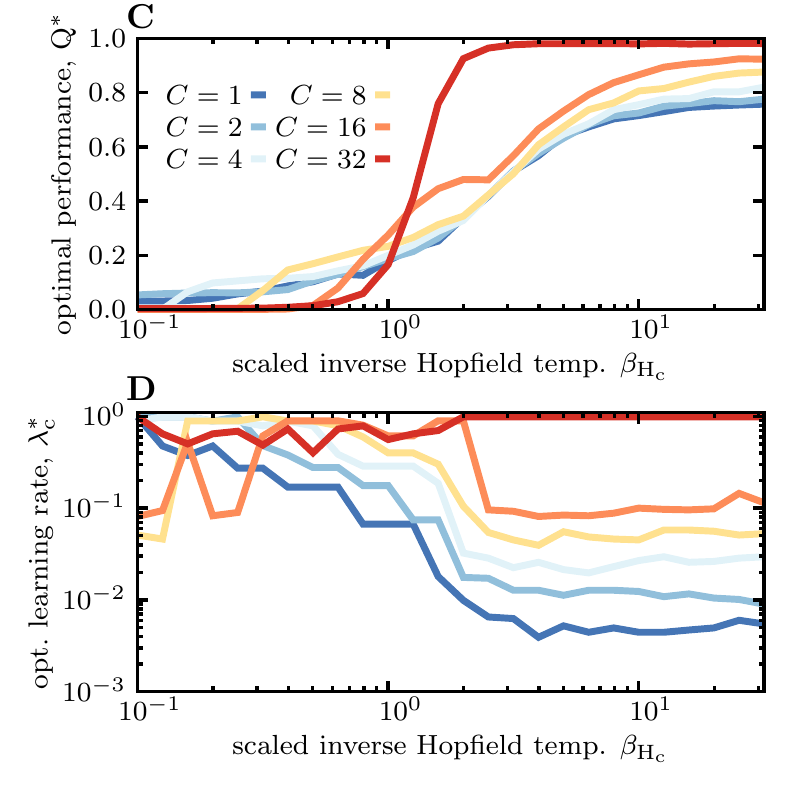}
 \caption{
\textbf{Optimal performance and learning rate at difference Hopfield  temperatures.} \textbf{(A)} The optimal accuracy of compartmentalized networks (i.e., for $\beta_\rS  \gg 1$) is shown  as a function of the scaled inverse Hopfield temperature $\beta_{\rH_\C} $ for difference number of compartments $C$ (colors) for static patterns ($\mu_\eff =0$); see Fig.~\ref{Fig4}B top. \textbf{(B)}  The optimal learning rate $\lambda^*_\C$ for each strategy as a function of $\beta_{\rH_\C} $ is shown. In contrast to Fig.~\ref{Fig3} the learning rate is not rescaled here and does not collapse for $ \beta_{\rH_\C}  \gg 1$.  As the equilibration noise increases (decreasing $\beta_{\rH_\C} $), networks with distributed memory ($C<N$) show two-step transitions (A). The first transition occurs at $\beta_{\rH_\C} \simeq 1$, which results in the reduced accuracy of the networks' memory retrieval. At this transition point, the networks' learning rate $\lambda_\C$ approaches its maximum value $1$ (B). Consequently, memory is only  stored for $C <N$ patterns (i.e., one pattern per sub-network) and the optimal performance $\Q^*$ is reduced to approximately $\frac{C}{N}$ (A). The second transition occurs at $\beta_{\rH_\C}  \approx \frac{1}{ N_\C} = \frac{C}{N}$, below which no pattern can be retrieved  and the performance approaches zero (A). (\textbf{C-D}) Similar to (A-B) but for evolving patterns with the effective mutation rate $\mu_\eff = 0.01$ similar to Fig.~\ref{Fig4}D. The number of presented patterns is set to N = 32. Similar to Figs.~\ref{Fig3},~\ref{Fig4} we keep $L \cdot C = \text{const.}$, with $L = 800$ used for networks with $C = 1$. }
\label{fig:Phasetransition_Beta_H_SI} 
\end{figure}

%%%%%%%%%%%%%%%%%%%%%%%%%%%%%%%%%%%%%%%%%%%%%%%%%%%%%%%
\end{document}